\documentclass[showpacs,12pt,amsmath,amssymb,aps,onecolumn]{revtex4-1}
\usepackage{graphics}
\usepackage{graphicx}
\usepackage{subfigure}
\usepackage{txfonts}
\usepackage{color}
\usepackage{bm}   
\usepackage{amsmath,mathrsfs}
\usepackage{CJK}
\usepackage{amssymb}
\usepackage{mathrsfs}
\usepackage{multirow}
\begin{document}

\title{Coherent photoproduction of $J/\psi$ and $\varUpsilon$ mesons in pp and PbPb ultraperipheral collisions from dynamical gluon distributions\thanks{This work is partly supported by the Key Research Program of Frontier Sciences, CAS, under the Grants Number NO. QYZDY-SSW-SLH006 of Chinese Academy of Sciences.}}


\author{%
      Qiang Fu()$^{1,2,3}$\email{fuqiang@impcas.ac.cn}%
\quad Xurong Chen()$^{1}$\email{xchen@impcas.ac.cn}%
}

\affiliation{
$^1$ Institute of Modern Physics, Chinese Academy of Sciences, Lanzhou, 730000, China\\
$^2$ Lanzhou University, Lanzhou 730000, China\\
$^3$ University of Chinese Academy of Sciences, Beijing 100049, China \\
}
\date{\today}

\begin{abstract}
We present calculations of coherent photoproduction of vector mesons ($J/\Psi$ and $\Upsilon$) with leading-order parton distribution functions to check new kinds of corrections of the DGLAP equations and nuclear modifications. The input gluon distribution of the proton is the dynamical parton model from the DGLAP equations with GLR-MQ-ZRS (Gribov-Levin-Ryskin, Mueller-Qiu, Zhu-Ruan-Shen) modifications. From comparison between several other gluon distribution models, we find that the dynamical gluon distribution fits with the results of meson photoproduction experiments in the high energy region. The calculation of the differential cross sections using dynamical and other gluon distributions is compared with the experimental data from the HERA, ZEUS and LHCb Collaborations. Although there is little data for the rapidity distribution of vector meson photoproduction near zero rapidity, the dynamical gluon distribution works well with the data in the large rapidity region.
\end{abstract}

\maketitle

\section{Introduction}
\label{SecI}
Measurements of vector meson photoproduction in coherent pp or nuclear collisions have been used as a way of determining the gluon densities in protons and nuclei. The correlation between the vector meson photoproduction cross sections and the gluon densities in the leading-order approximation also gives us a unique opportunity to test the gluon distribution function over a large range of Bjorken-x, especially the pp collisions at $\sqrt{s_{NN}}~=~7~\rm{TeV}$ from the LHCb Collaboration~\cite{0954-3899-41-5-055002,Aaij2015}. For a long time, PDF set analyses have achieved $\alpha_{s}$ next-leading-order and even next-to-next-leading-order results. Besides higher order corrections of the DGLAP equations, we can also consider higher twists of the partons' evolution. Below, we calculate the diffractive exclusive photoproduction of vector mesons by a dynamical gluon distribution from GLR-MQ-ZRS (Gribov-Levin-Ryskin, Mueller-Qiu, Zhu-Ruan-Shen) corrections to the Dokshitzer-Gribov-Lipatov-Altarelli-Parisi (DGLAP) equations~\cite{doi:10.1142/S0218301314500578}. These corrections involve higher twist Feynman diagrams describing gluon recombination effects. The method of investigating gluon density by vector meson photoproduction was proposed in Ref.~\cite{PhysRevC.65.054905}. This method can be used to check the rationality of a gluon distribution function from experimental data on production of vector mesons. By using our new gluon distribution to reproduce some special experimental data, we hope to test the corrections to the DGLAP equations with ultraperipheral meson photoproduction.

For ultraperipheral collisions, the photon flux model given by the Weizsacker-Williams method~\cite{Fermi1925} with impact parameter larger than $2R$ is essential for calculating vector meson photoproduction. The virtual photon flux at a distance r away from a charge Z nucleus is
\begin{eqnarray}
\begin{aligned}
\frac{d^{3}N_{\gamma}} {dkd^{2}r}=\frac{Z^{2}\alpha _{em}\xi^{2}} {\pi^{2}kr^{2}} \left[K_{1}^{2}(\xi)+\frac{1}{\gamma_{L}^{2}}K_{0}^{2}(\xi) \right]
\label{eq:1_1}
\end{aligned}
\end{eqnarray}
where $\xi=kr/ \gamma_{L}$, $K_{0}(\xi)$ and $k_{1}(\xi)$ are modified Bessel functions, which give reasonable truncation of the photon energy due to the radius of the nucleus. At the Large Hadron Collider (LHC), the Lorentz term $\gamma_{L}~=~\sqrt{s_{NN}}/2m$ for $\sqrt{s_{NN}}~=~2.76~\rm{TeV}$ PbPb collisions is 1470, or 7455 for pp collisions. The total photon flux hitting the target nucleus is the integral of Eq.~(\ref{eq:1_1}) over the transverse distance of the impact parameter when the two nuclei do not overlap. A reasonable analytic approximation~\cite{BERTULANI1988299} for nucleus AA collisions is given by the photon flux integrated over the distance r larger than $2R_{A}$~\cite{BERTULANI1988299,PhysRevD.42.3690,BAUR1990786}.
\begin{eqnarray}
\begin{aligned}
\frac{dN_{\gamma}}{dk}=\frac{2Z^{2}\alpha _{em}}{\pi k}\times \left[\xi_{R}^{AA}K_{0}(\xi_{R}^{AA})K_{1}(\xi_{R}^{AA})-\frac{(\xi_{R}^{AA})^{2}}{2} \left(K_{1}^{2}(\xi_{R}^{AA})-K_{0}^{2}(\xi_{R}^{AA}) \right) \right]
\label{eq:1_2}
\end{aligned}
\end{eqnarray}
where $\xi_{R}^{AA} = 2kR_{A}/\gamma_{L}$, $R_{A} = 1.2A^{1/3} [\rm{fm}]$. The elastic photon differential flux with photon energy is ~\cite{Ryskin1997}
\begin{eqnarray}
\begin{aligned}
\frac{dN_{\gamma/p}}{dk}=\frac{\alpha _{em}}{2\pi k}\left[ 1+\left( 1-\frac{2k}{ \sqrt{ s_{NN} } } \right)^2 \right]\times \left( \rm{ln}~\Omega-\frac{11}{6}+\frac{3}{\Omega}-\frac{3}{2\Omega^2}+\frac{1}{3\Omega^3} \right)
\label{eq:1_3}
\end{aligned}
\end{eqnarray}
with $\Omega = 1+[(0.71~\rm{GeV}^2)/Q_{min}^2]$ and $Q_{min}^2 = k^2/[\gamma_L^2(1-2k/\sqrt{s_{NN}})]$. After determining the total flux of  photons, in principle, we can easily calculate any photoproduction's total cross section by just factorizing these processes into the photonucleus and the photon emission flux. Thus we can get the cross section of coherent vector meson photoproduction.
The photonuclear-induced vector meson photoproduction described in this work depends on the gluon distribution functions in the nucleon or nucleus. Usually, a gluon PDF at certain $Q^2$ is achieved by evolving the parameterized PDFs using the DGLAP equations at an initial $Q_{0}^2$, at which $Q_{0}^2$ the initial distributions of gluon and sea quarks are $not$ assumed to be zero. In contrast to mainstream gluon PDFs~\cite{PhysRevC.87.027901, PhysRevLett.92.142003, PhysRevC.84.024916, Goncalves2016} used in the calculation of vector meson photoproduction, we use a gluon distribution function of dynamic evolution with non-linear corrections including gluon-gluon recombination and gluon absorption by quarks~\cite{WEIZHU1999245, ZHU1999378, Zhu:2004xj}. Furthermore, we assume zero gluon and sea quark distributions at a very low initial $\rm{Q_{0}^2 = 0.064~GeV^2}$ and all the gluons and sea quarks  generated by quark or gluon radiation. In other words, this new PDF set~\cite{doi:10.1142/S0218301314500578} considers not only gluon generation from quark and gluon radiation but also some mechanisms for limiting the unrestrained increasing of gluons. 

\section{Dynamical Parton Distributions}
\label{SecII}
The parton distributions of nucleons reveal the inner structure of the nucleons and information about their QCD evolution. In high energy photon scattering, the gluon density dominates the partons' evolution in the nucleon, which is critical in understanding the appearance of nonlinear behavior~\cite{doi:10.1146/annurev.nucl.010909.083629}. Currently, most of the available parameterizations of parton distributions were achieved by global analysis of the experimental data with the linear QCD evolution equations: the DGLAP equations. The solutions of the DGLAP equations depend on the initial parameterized distributions at the low starting scale $Q_0^2~\sim~1~\rm{GeV}^2$, commonly called the non-perturbative input, which uses complicated functions with many free parameters. The current methods of determining the PDFs take a starting point fixed at an arbitrary $Q_0^2~\sim~1~\rm{GeV}^2$ and the input parton distributions are parameterized by comparing the evolved initial PDFs with experimental data at the same $Q^2$s. These input distributions are irrelevant to any physical models with reasonable assumptions. Some models even input negative gluon distributions at $Q_0^2$. Although a lot of progress has been made, the gluon distribution at small x still has quite large uncertainties~\cite{PhysRevC.84.024916, Goncalves2016}.

\begin{figure}[htp]
\centering
{
\begin{minipage}[b]{0.55\textwidth}
\includegraphics[width=1\textwidth]{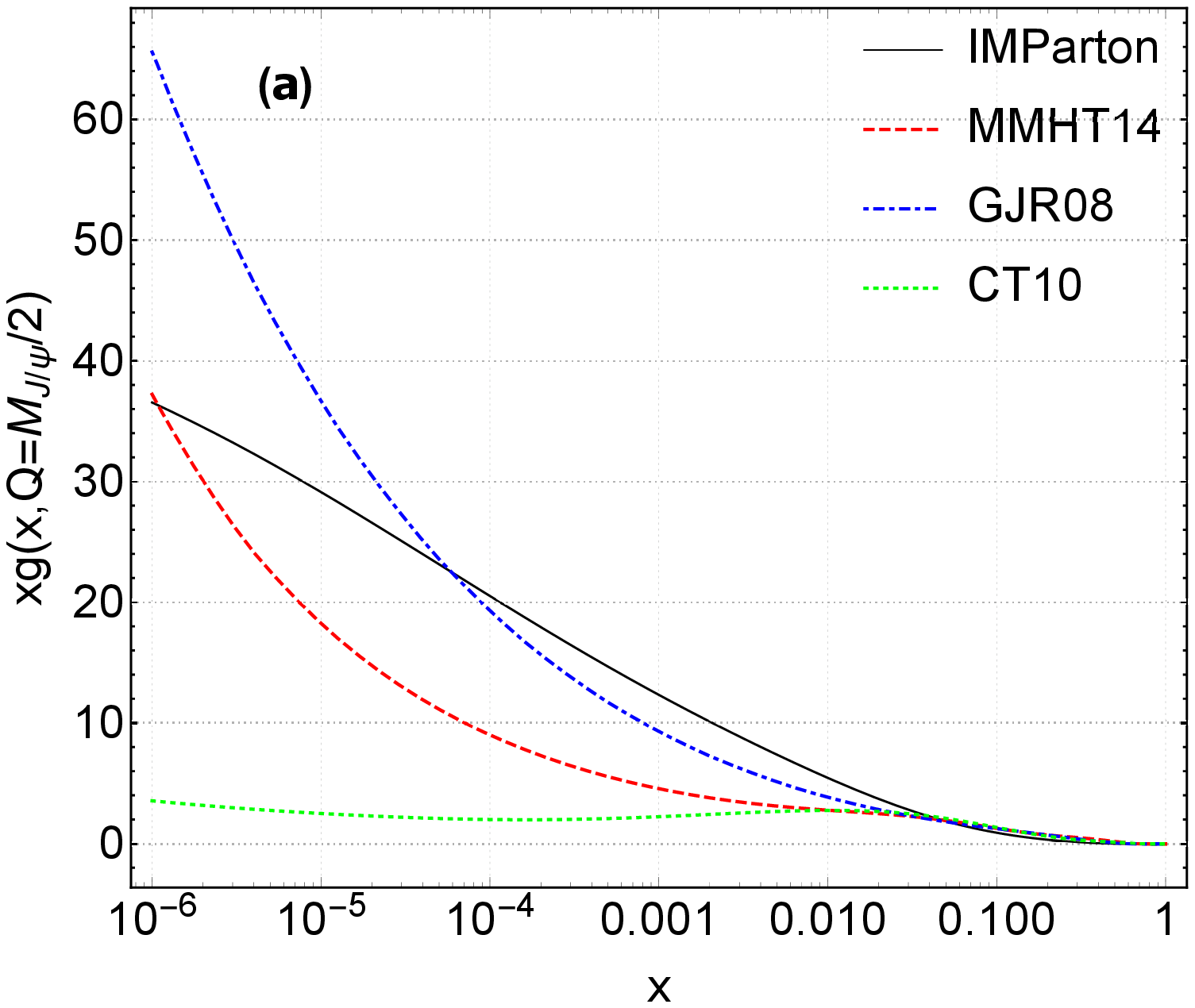}
\end{minipage}
}
{
\begin{minipage}[b]{0.55\textwidth}
\includegraphics[width=1\textwidth]{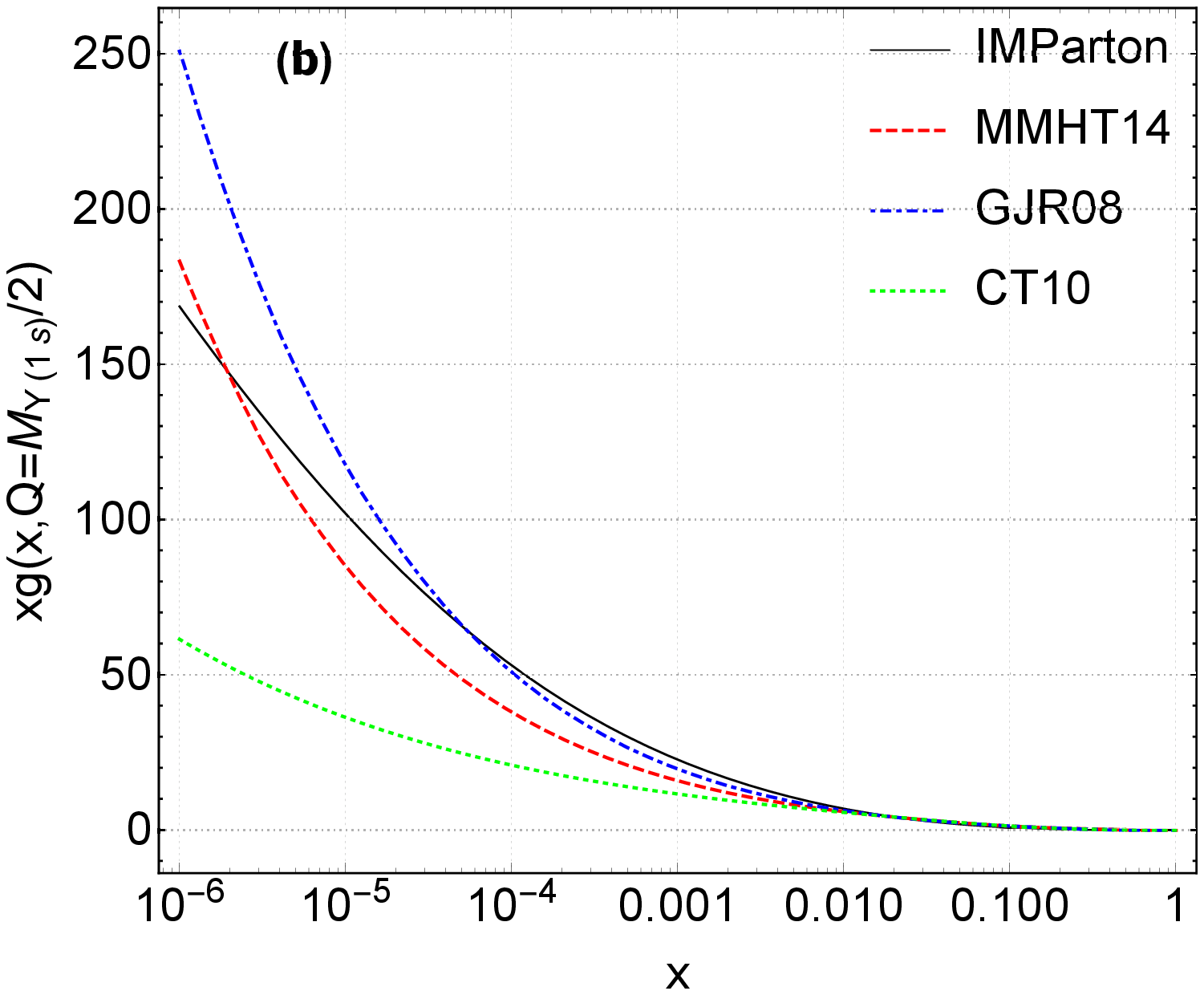}
\end{minipage}
}
\caption{(color online). Gluon distributions from different groups~\cite{1674-1137-41-5-053103, Harland-Lang2015, Gluck2008, PhysRevD.82.074024} with leading-order approximation: (a) at $Q~=~M_{J/\Psi}/2$, corresponding to $J/\Psi$ photoproduction and (b) at $Q~=~M_{\Upsilon}/2$, corresponding to $\Upsilon$ photoproduction.}
\label{Fig:2_1}
\end{figure}

Although the constituents of a hadron are complex, at very high energies the photohadron scattering processes are dominated by gluon densities. Thus, the determination of gluon distribution in the proton can be improved by the many available experimental data from ep collisions. However, the behavior of the gluon distribution in the small x region still has a large uncertainty, as seen in Fig~\ref{Fig:2_1}. In this paper, we tried the dynamical parton evolution approach~\cite{doi:10.1142/S0218301314500578, 1674-1137-41-5-053103}, in which the initial parton distributions are inspired by the simple quark model. The initial parton distributions consist only of three valence quarks at initial $Q^2_0$. There is zero gluon density at the input scale. All sea quarks and gluons in the small x region are purely dynamically generated by radiation in the DGLAP equations with nonlinear corrections (gluon-gluon or gluon-quark recombinations). The dynamical parton model is developed and extended to a very low scale $Q_0^2~=~0.064$~GeV$^2$ in Ref.~\cite{doi:10.1142/S0218301314500578}.

\section{Elastic Photoproduction of Vector Mesons}
\label{SecIII}

\subsection{$\gamma p \to Vp$ process}
In this paper, we associate the gluon distribution function with the vector meson photoproduction amplitude using a two-gluon exchange model in pertubative QCD~\cite{PhysRevD.50.3134}. It is obvious that higher-order corrections of the gluon density are important. However, full inclusion of higher-order effects still remains a great challenge. We assume that the corrections of higher-order effects is a normalization of the value of the $\gamma p \to Vp$ cross section, i.e. the behavior of the cross section remains the same, except for the scale of its value. Thus we use a phenomenological multiplicative correction parameter $\xi_{V}$ to include other effects such as next-to-leading-order effects~\cite{PhysRevD.57.512}. For forward-scattering (near $t~=~0$) of elastic photoproduction on a proton, the corrected leading-order amplitude is expressed as
\begin{eqnarray}
\begin{aligned}
\frac{d\sigma^{\gamma p\to Vp}}{dt}\left.\right| _{t=0} = \xi_{V}\left(\frac{16\pi^{3}{\alpha_s}^2\Gamma_{ee}}{3\alpha_{em}{M_v}^5}\right)\left[xg_{p}(x,Q^2)\right]^2 ,
\label{eq:3_1}
\end{aligned}
\end{eqnarray}
where $M_{V}$ stands for the mass of the generated vector meson, $x = M_{v}^2/W_{\gamma p}^2$ is the Bjorken-x that represents the fraction of the proton momentum carried by the parton, and $g_{p}(x, Q^2)$ is the gluon distribution function, which is evaluated at the momentum transfer squared $Q^2 = (M_{v}/2)^2$. Equation~(\ref{eq:3_1}) can easily be extended to the nuclear target:
\begin{eqnarray}
\begin{aligned}
\frac{d\sigma^{\gamma A\to VA}}{dt}\left.\right| _{t=0} = \xi_{V}\left(\frac{16\pi^{3}{\alpha_s}^2\Gamma_{ee}}{3\alpha_{em}{M_v}^5}\right)\left[xG_{A}(x,Q^2)\right]^2
\label{eq:3_2}
\end{aligned}
\end{eqnarray}
where $G_{A}(x,Q^2) = g_{p}(x,Q^2)\times R_{g}^A(x, Q^2)$ is the nuclear gluon distribution function and $R_{g}^A(x, Q^2)$ is the nuclear modification of gluon distribution. In this work, we use the EPS09~\cite{1126-6708-2009-04-065} nPDF nuclear modification for the PDFs by linear DGLAP. For our IMParton, we exploit the nIMParton~\cite{Wang20171} which is compatible with the IMParton PDFs by DGLAP equations with nonlinear corrections. 

\begin{figure}[htp]
\centering
{
\begin{minipage}[b]{0.55\textwidth}
\includegraphics[width=1\textwidth]{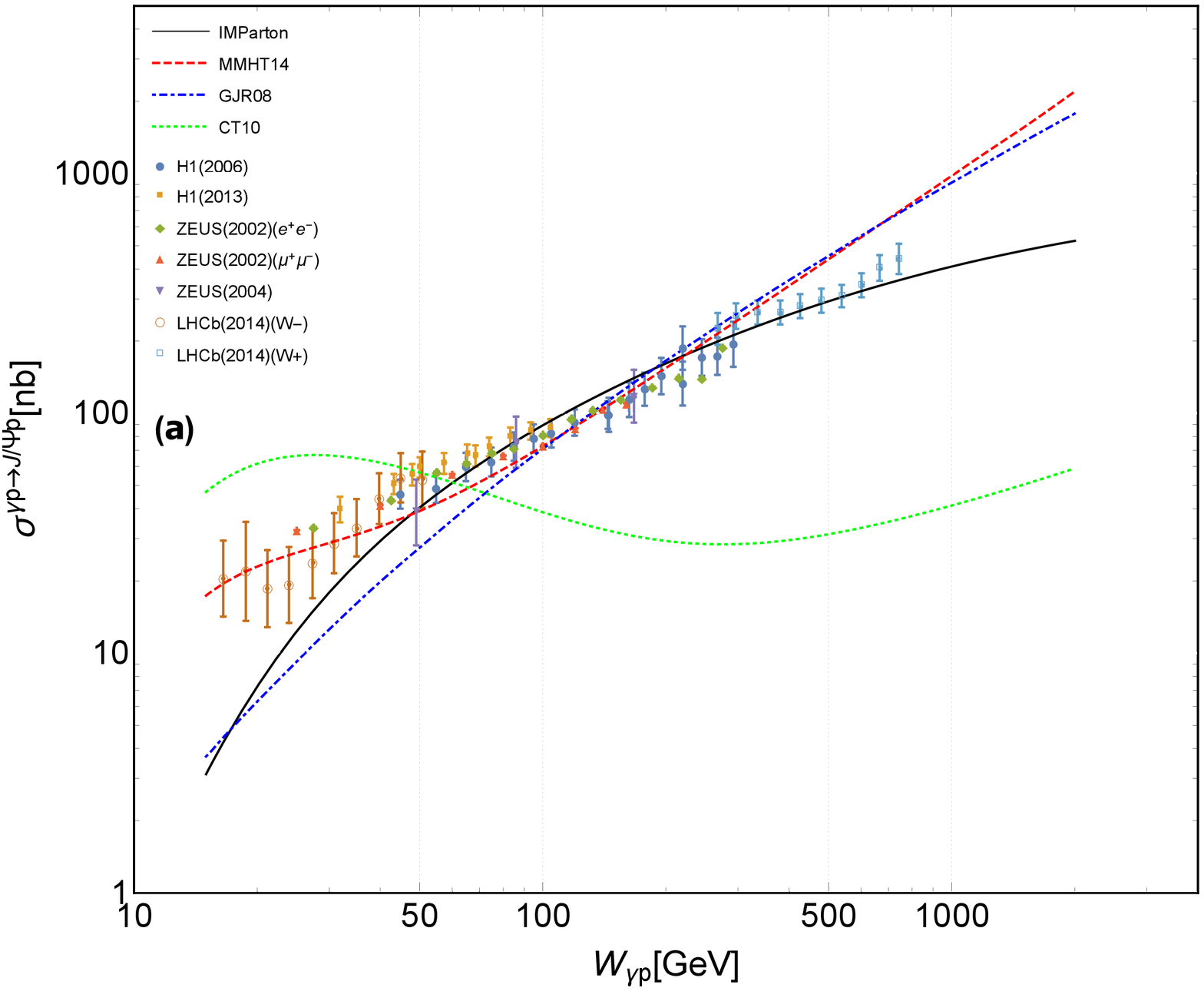}
\end{minipage}
}
{
\begin{minipage}[b]{0.55\textwidth}
\includegraphics[width=1\textwidth]{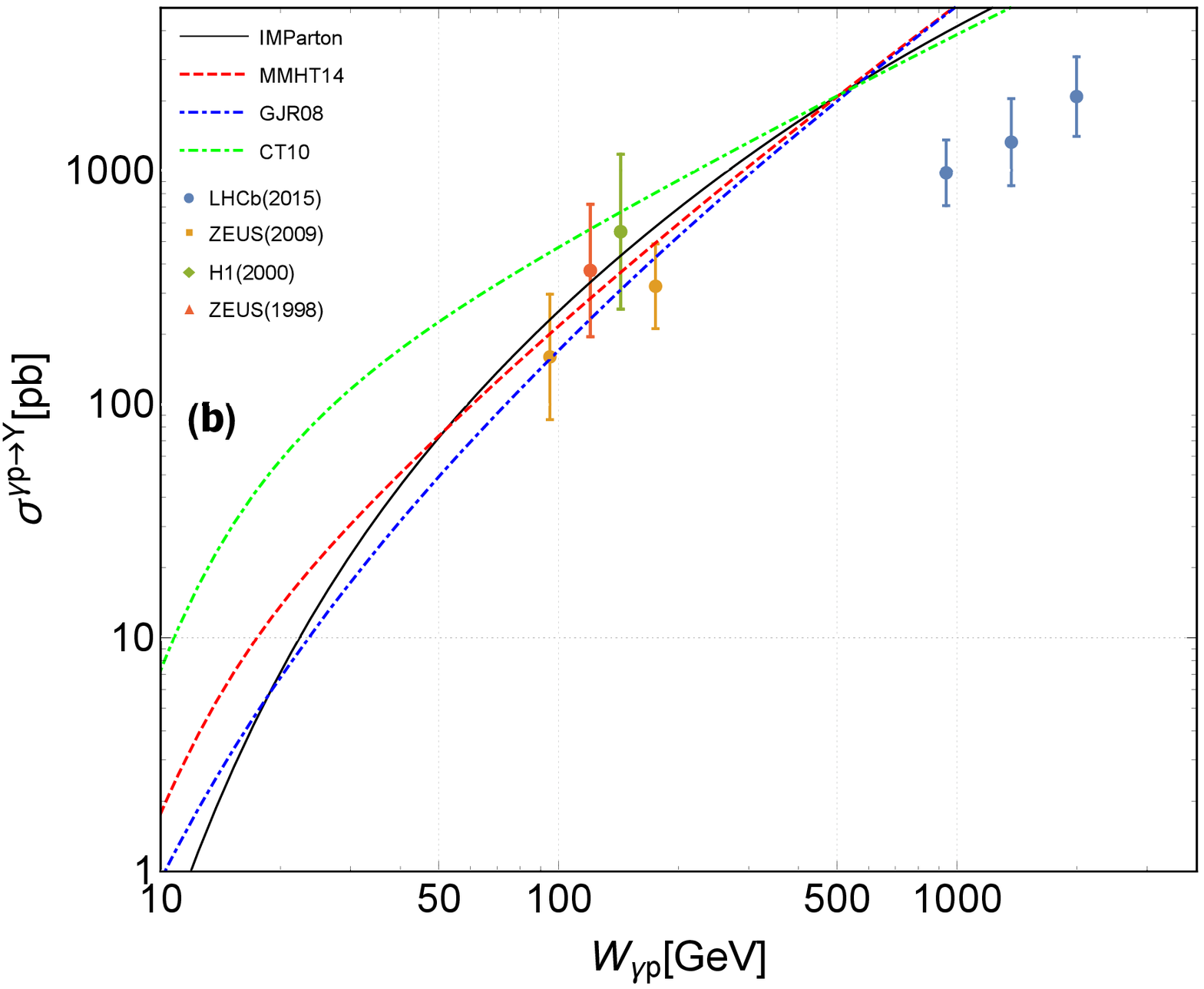}
\end{minipage}
}
\caption{(color online). Meson photoproduction cross sections for different gluon distribution models: $\gamma p$ center-of-mass energy dependence of the diffractive (a) $J/\Psi$ and (b) $\Upsilon$ photoproduction cross sections reproduced with the LO gluon distribution functions with one free parameter. The solid black curve is the corrected LO IMParton result fitted with data from~\cite{Chekanovetal.2002, Chekanov20043, Alexa2013, 0954-3899-41-5-055002}~(\cite{Aaij2015, Adloff200023, Breitweg1998432, Chekanov20094}). The other curves are cross sections from different LO gluon distributions, namely MMHT14, GJR08 and CT10.}
\label{Fig:3_1}
\end{figure}

The correction factor $\xi_V$ is approached by fitting the calculated cross sections of the elastic photoproduction of vector mesons on protons, $\sigma^{\gamma p\to Vp}(W_{\gamma p})$, to reproduce the experimental data from ZEUS and HERA from Ref.~\cite{Alexa2013, Harland-Lang2015, Chekanovetal.2002, Chekanov20043} and LHCb from Ref.~\cite{0954-3899-41-5-055002}. Considering the t dependence of $d\sigma^{\gamma p\to J/\Psi p}/dt$, we note that its $p_t$ dependence can be parameterized in the exponential form of $\rm{exp[-b(W_{\gamma p})p_t^2]}$, in which the slope parameter $b_{J/\Psi}$ weakly depends on the $\gamma p$ center-of-mass energy according to the Regge-motivated parameterization. We take a slope parameter
\begin{eqnarray}
\begin{aligned}
b = 4.5+\rm{ln}(\frac{W_{\gamma p}}{90~[\rm{GeV}]})
\label{eq:3_3}
\end{aligned}
\end{eqnarray}
 as the transverse size of the charge radius of a proton to obtain the total cross section $\sigma^{\gamma p\to Vp} = 1/b~d\sigma^{\gamma p\to Vp}\left.\right|_{t=0}$. We use this parameter to express the exponential decrease of the differential cross sections with four momentum transfer t. We get $\xi_{V}$ results, see Table~\ref{fit}, by fitting the elastic photoproduction of HERA data from newer analyses, ZEUS and LHCb~\cite{PhysRevLett.113.232504, Adloff200023, 0954-3899-40-4-045001, 0954-3899-41-5-055002, Harland-Lang2015, Chekanovetal.2002, Chekanov20043}, and the results are presented in Fig.~\ref{Fig:3_1} together with predictions.

The photoproduction cross section is obtained from Eq.~\ref{eq:3_1} assuming $\overline{Q} = M_{v}/2$. We can see that not all the different models for the gluon distributions describe the experimental data similarly. CT10 decreases in the range of $10^{-4}<x_{B}<10^{-2}$, which lead to anomalous behavior in $\sigma^{\gamma p\to J/\Psi p}$ in Fig.~\ref{Fig:3_1}(a). MMHT14 matches well at lower energy, while it increases a little bit fast at higher energy. GJR08 is compatible in the range of 100 GeV to 300 GeV of $\gamma p$ center - of - mass energies. IMParton's gluon distribution is generated from pure dynamical gluons and this model considers gluon recombinations, which give a depression of gluon density at small x and leads to a reasonable result at higher energy range. However, the fewer parameters of the IMParton PDF set makes it hard to describe lower energies very well. Cross sections from LO MMHT14 and GJR08 reproduce the low energy range very well since their evolution starting points are close to the lower limits of the experiments' $Q^2$. Our IMParton PDF starts at very low $Q^2~=~0.064~\rm{GeV^2}$. Moreover, as the difference between the CT10 distribution and other models increases at low energy, the prediction for the rapidity distributions does not conform to the LHCb data.

In Table \ref{fit} we list the fitted parameters for different gluon distribution models and compare the predictions. The data from HERA, ZEUS and LHCb are also shown in Fig.~\ref{Fig:3_1}. The results give reasonable values of $\chi^2/ndf$ for the IMParton parameterization for $J/\Psi$ production, which describes the data well in the high energy region. As can be seen from Fig.~\ref{Fig:3_1}, the predictions of different parameterizations diverge  in the region of higher than HERA data. What should be emphasized is the fitting result for $\xi$ obtained by using the IMParton parameterization. It indicates that in $J/\Psi$ photoproduction calculations with LO two-gluon exchange, a gluon recombination effect in the small x region is necessary.

\begin{table}[htp]
\centering
\caption{Values of the free parameters $\xi_{J/\Psi(\Upsilon(1s))}$ for different gluon parameterizations obtained by the minimization of $\chi^2/ndf$. The $\chi^{2}/ndf$ values are shown for comparison.}
\footnotesize
\begin{tabular*}{100mm}{c@{\extracolsep{\fill}}ccccc}
\toprule Models & $\xi_{J/\Psi}$   & $\chi^2(J/\Psi)$ & $\xi_{\Upsilon(1s)}$   & $\chi^2(\Upsilon(1s))$ \\
\hline
IMParton\hphantom{0} & \hphantom{0}1/15.2 & \hphantom{0}1.38 & \hphantom{0}1/16.2 & \hphantom{0}1.51 \\
MMHT14\hphantom{0} & \hphantom{0}1/2.6 & \hphantom{0}3.04 & \hphantom{0}1/9.3 & \hphantom{0}1.67\\
GJR08\hphantom{0}  & \hphantom{0}1/10.9 & \hphantom{0}5.74 & \hphantom{0}1/17.1 & \hphantom{0}2.04\\
CT10\hphantom{00}  & \hphantom{0}1/1.1 & \hphantom{0}24.74 & \hphantom{0}1/2.4 & \hphantom{0}0.40\\
\hline
\end{tabular*}
\vspace{0mm}
\label{fit}
\end{table}

\subsection{Nuclear modification of elastic photoproduction of vector mesons}
The differential cross section of coherent vector meson photoproduction on a nucleus can be factorized into the forward limited amplitude $d\sigma^{\gamma A\to VA}/dt\left.\right |_{t=0}$ (see Eq.~\ref{eq:3_2}) and the form factor squared, $\left|F(t)\right|^2$, of the nucleus. The former term encodes the dynamical information of the photo-nucleus interaction while the latter determines the momentum transfer during diffractive scattering. The nuclear form factor can be treated as the momentum space analog of the nuclear density distribution, which derived from the Fourier transform of the spatial density distribution, $\int d^3r\rho(r)e^{i\bm{q}\bm{r}}$, where q is the four-momentum transfer squared of the nuclei. It is customary to use the Woods-Saxon distribution, $\rho(r) = \frac{\rho_{0}}{1+\rm{Exp}\left[(r-R_{A})/d\right]}$, for a spherically symmetric heavy nuclei with central density $\rho_{0}$, radius $R_{A}$ and nuclear neutron skin thickness d. For the $^{208}\rm{Pb}$ nuclei used by the LHCb experiment, $\rho_{0}~=~0.16/\rm{fm}^{3}$, $R_{A}~=~1.2A^{1/3}~\rm{fm}$ and $d~=~0.549~\rm{fm}$~\cite{DEVRIES1987495}. However, it is hard to get the analytic form of the Fourier transform of the Woods-Saxon distribution, so we turn to using an approximation of $\rho(r)$ to calculate $F(t)$ with the form of convolution of a well-distributed function with a Yukawa distribution~\cite{BERTULANI2000527, PhysRevC.14.1977, PhysRevC.65.054905}. Thus we have the approximate form factor 
\begin{eqnarray}
\begin{aligned}
F(q = \sqrt{|t|}) = \frac{4\pi\rho_0}{Aq^3}\left[sin(qR_A)-qR_Acos(qR_A)\right]\left[\frac{1}{1+a^2q^2}\right] ,
\label{eq:3_4}
\end{aligned}
\end{eqnarray}
where $a~=~0.7~\rm{fm}$ is the upper limit of the Yukawa potential radius, while the form factor is just the convolution of the Fourier transform of an even density sphere and Yukawa distribution. The cross section of photoproduction of vector mesons on nuclei is then written as
\begin{eqnarray}
\begin{aligned}
\sigma^{\gamma A\to VA}(k) = \frac{d\sigma^{\gamma A\to VA}}{dt}\left.\right|_{t=0} = \int_{t_{min}(k)}^{\infty}dt|F(t)|^2 ,
\label{eq:3_5}
\end{aligned}
\end{eqnarray}
where $t_{min}(k) = (M_V^2/4k\gamma_L)^2$ is a proper cut on momentum transfer for a narrow resonance~\cite{PhysRevLett.92.142003}. Hence the differential cross section of coherent vector meson production is a product of the nucleus photoproduction cross section with the photon flux
\begin{eqnarray}
\begin{aligned}
\sigma^{A[\gamma]A\to AAV}(k) &= k\frac{dN_{\gamma}(k)}{dk}\times \int_{t_{min}(k)}^{\infty}dt\frac{d\sigma{\gamma A\to VA}}{dt}\left.\right|_{t=0}|F(t)|^2
\label{eq:3_6}
\end{aligned}
\end{eqnarray}
Practically, we usually represent the cross section of heavy ion collisions by the rapidity of the produced particles. The photon energies in the laboratory frame k are related to the rapidity y in the form of $k = (M_V/2)exp(y)$. Under this substitution, the differential cross section with the rapidity is thus written as
\begin{eqnarray}
\begin{aligned}
\frac{d\sigma^{AA\to AAV}}{dy} &= \left[k\frac{dN_{\gamma}(k)}{dk}\sigma^{\gamma A\to VA}(k)\right]_{k = k_l} + \left[k\frac{dN_{\gamma}(k)}{dk}\sigma^{\gamma A\to VA}(k)\right]_{k = k_r}
\label{eq:3_7}
\end{aligned}
\end{eqnarray}
The rapidity distribution is obviously symmetric with zero rapidity, since the symmetric collision energies and same nuclei for both projectile and target make the left and right photon flux identical in both directions~\cite{Thomas:2016oms}. 

\section{$J/\Psi$ and $\Upsilon$ production in pp (PbPb) collision at $\sqrt{s_{NN}}~=~7~(2.76)~\rm{TeV}$}
\label{SecIV}
Different from free nucleons, the PDFs of a nucleus are influenced by nuclear medium effects. The study of quarks and gluon distributions in a variety of nuclei has been a hot issue since the discovery of the EMC effect in the 1980s~\cite{AUBERT1983275}. It is also one of the physics goals of the LHC, whose four major experiments are able to detect coherent photoproduction of vector mesons in ultraperipheral PbPb collisions. In particular, the results of $J/\Psi$ photoproduction at $\sqrt{s_{NN}}~=~2.76~\rm{TeV}$ PbPb collision have been released by the ALICE Collaboration~\cite{Abelev20131273, Abbas2013}. Not long afterwards, the LHCb Collaboration measured $J/\Psi$ and $\Upsilon(1s)$ photoproduction in pp collisions at $\sqrt{s_{NN}}~=~7~\rm{TeV}$~\cite{0954-3899-40-4-045001, 0954-3899-41-5-055002}.

\begin{figure}[htp]
\centering
{
\begin{minipage}[b]{0.55\textwidth}
\includegraphics[width=1\textwidth]{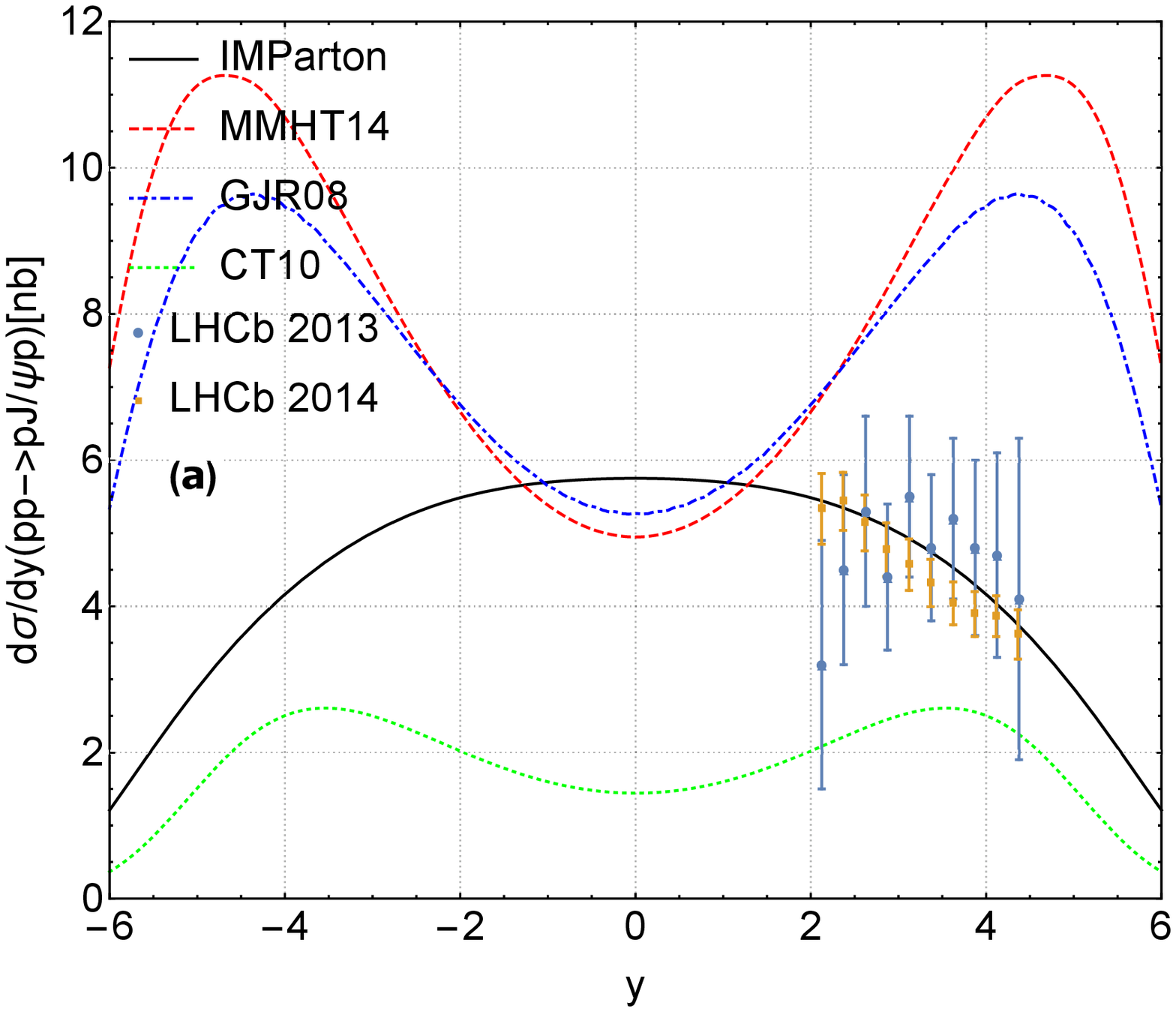}
\end{minipage}
}
{
\begin{minipage}[b]{0.55\textwidth}
\includegraphics[width=1\textwidth]{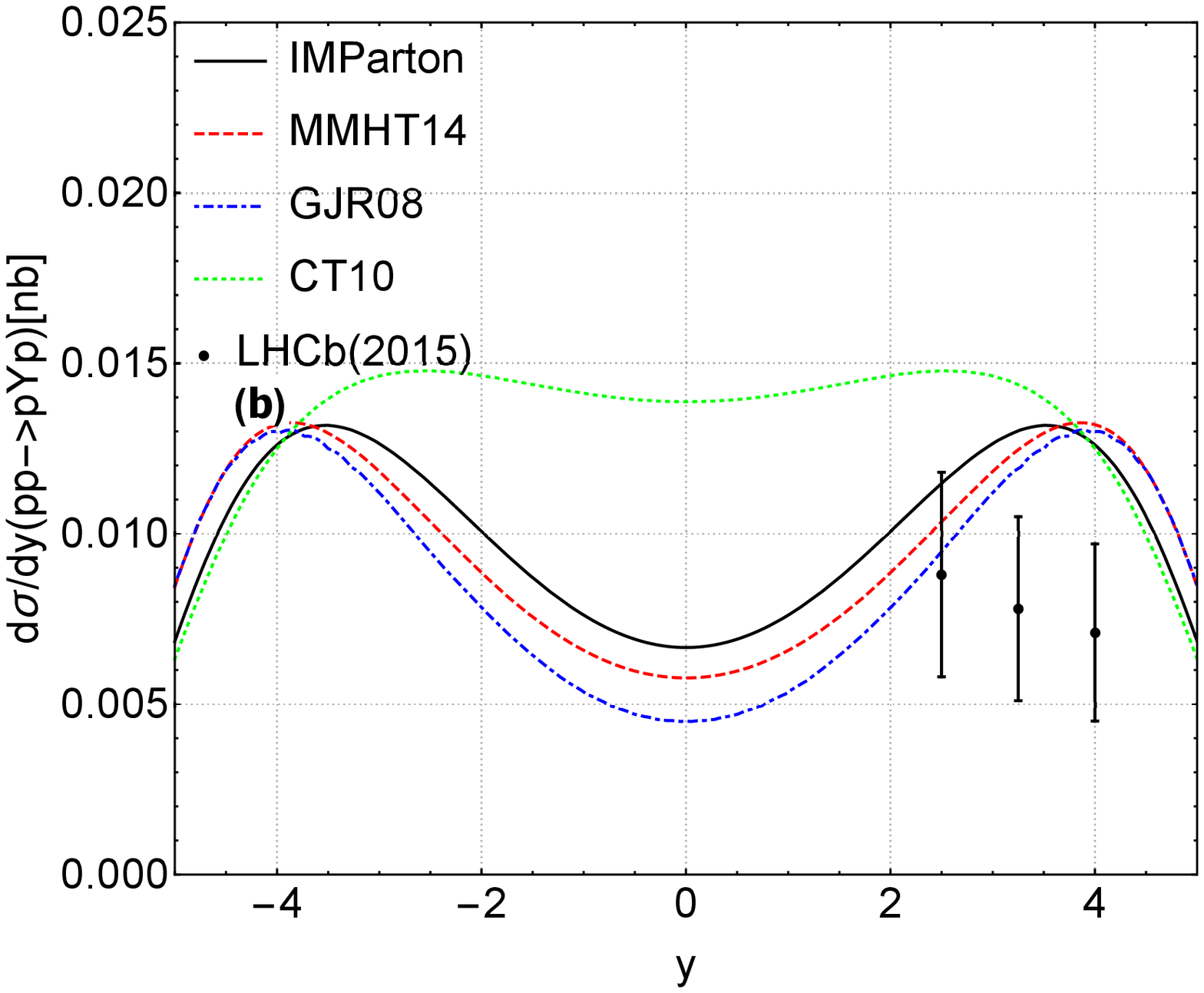}
\end{minipage}
}
\caption{(color online). Comparison of the rapidity distributions of the (a) $J\Psi$ and (b) $\Upsilon(1s)$ photoproduction cross sections in pp collisions at $\sqrt{s_{NN}}~=~7~\rm{TeV}$ predicted by different proton gluon distribution models. The data from Refs.~\cite{0954-3899-40-4-045001, 0954-3899-41-5-055002, Alexa2013, Chekanovetal.2002, Breitweg1998432, AID19963, Adloff200023, 0954-3899-41-5-055002} and Ref.~\cite{Aaij2015}, respectively, are also presented.}
\label{Fig:4_1}
\end{figure}

Previous experimental data from HERA and LHCb cover the $\gamma p$ center-of-mass energies from tens of GeV to nearly a thousand GeV, which means a large range of rapidity distribution of the generated mesons. We therefore present our predictions on $J/\Psi$ production from pp coherent scattering at the energy of $\sqrt{s_{NN}}~=~7~\rm{TeV}$ using the LO two-gluon exchange model in perturbative QCD. We also present the LHCb data~\cite{0954-3899-40-4-045001, 0954-3899-41-5-055002} together with the predictions from gluon distributions, as seen in Fig.~\ref{Fig:4_1}(a). Similarly, we give the corresponding $\Upsilon(1s)$ rapidity distribution predictions along with the LHCb data~\cite{Aaij2015} in Fig.~\ref{Fig:4_1}(b).

In Fig.~\ref{Fig:4_1}, we show the corresponding predictions for the rapidity distributions of cross section for $J/\Psi$ and $\Upsilon(1s)$ photoproduction in exclusive pp collisions at $\sqrt{s_{NN}}~=~7~\rm{TeV}$. We compare our prediction using IMParton with other gluon distributions and recent experimental data from the LHCb Collaboration~\cite{0954-3899-40-4-045001, 0954-3899-41-5-055002}. The MMHT14, GJR08 and CT10 models show a strong increase in the small x region, with double peaks in the rapidity distributions. The IMParton parameterization, on the other hand, predicts a plateau in the rapidity distribution at medium rapidity. This parameterization reproduces the LHCb data of the $J/\Psi$ production well. We should emphasize that some y distribution experimental data were extracted by the model proposed in Ref.~\cite{Jones2013}. We find that different gluon distributions do not all reproduce LHCb data in the same way, especially at medium rapidity. The future CMS experiment for central rapidity will probably give us more information.

\begin{figure}[htp]
\begin{center}
\includegraphics[width=0.55\textwidth]{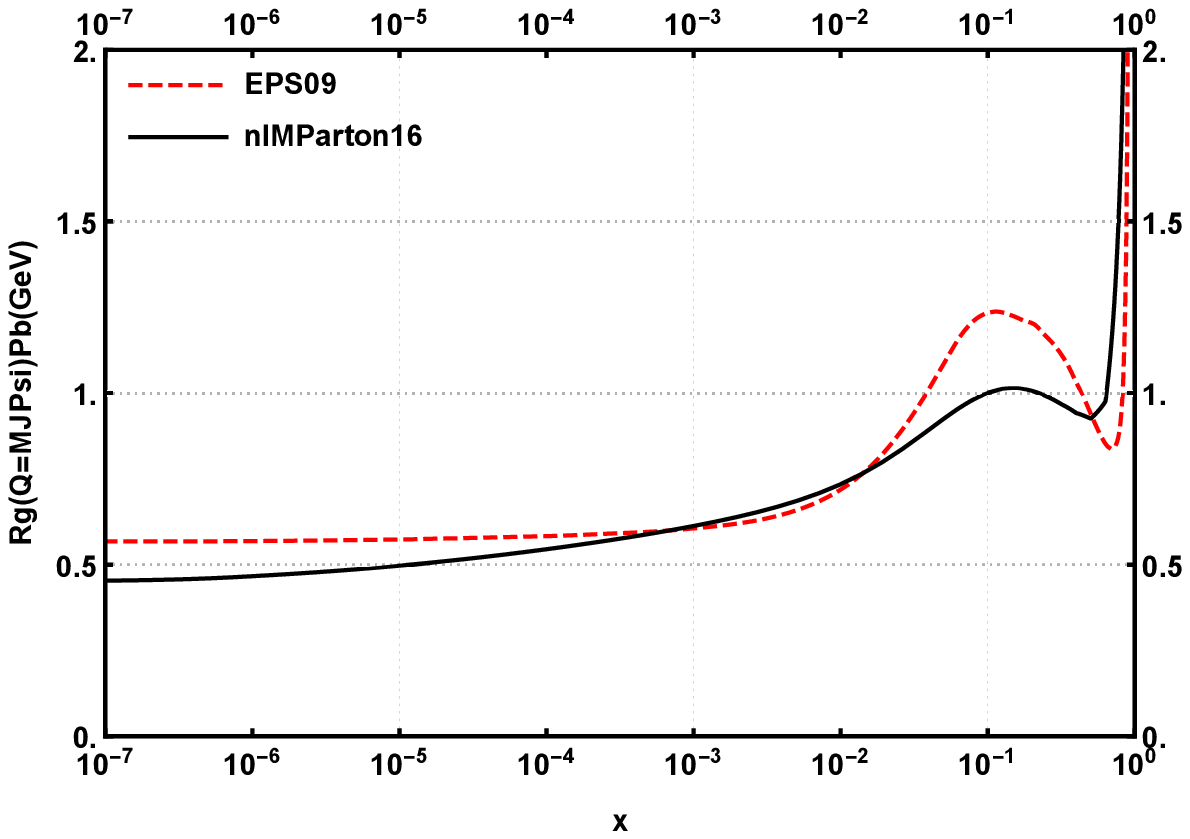}
\caption{(Color online) Nuclear medium modification, $R_{g}^{Pb}(x, Q^2=M_{J/\Psi}^2)$}, for gluon distribution in $^{208}\rm{Pb}$. The modification's $Q^2$ scale is at $M_{J/\Psi}^2$ (for the $J/\Psi$ photoproduction calculation). The red dashed curve stands for the EPS09 model, and the solid black curve represents the nIMParton16 model.
\label{Fig:4_2}
\end{center}
\end{figure}

\begin{figure}[htp]
\begin{center}
\includegraphics[width=0.55\textwidth]{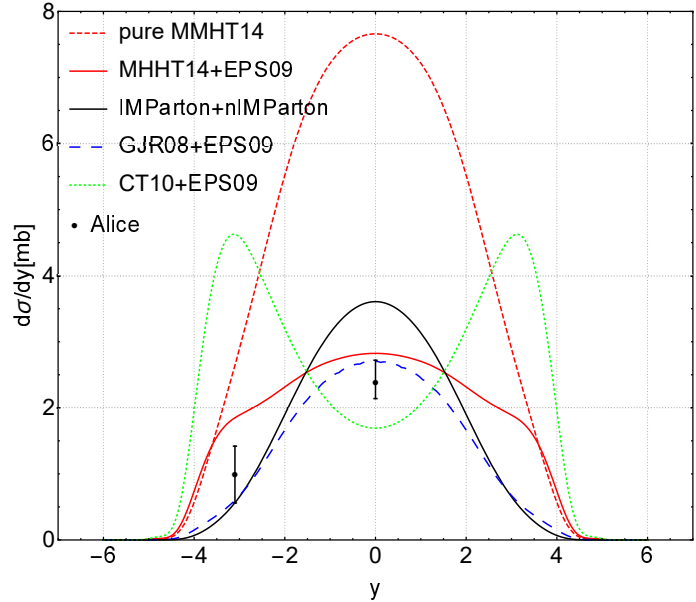}
\caption{(Color online) The rapidity distribution of the $J/\Psi$ photoproduction cross sections in PbPb ultraperipheral collisions at $\sqrt{s_{NN}}~=~2.76~\rm{TeV}$. The data from Ref.~\cite{Abelev20131273, Abbas2013} are also presented.}
\label{Fig:4_3}
\end{center}
\end{figure}

Since our IMParton PDFs involve nonlinear corrections, the nuclear modification called nIMParton16~\cite{Wang20171} is employed in the PbPb collisions. These nuclear medium modifications of parton distributions are from the global analysis of deeply inelastic scattering data of various nuclear targets. The nuclear dependence of nuclear modification is studied with only two free parameters. The comparison of nIMParton16 and EPS09 is shown in Fig.~\ref{Fig:4_2}. From this figure, we find that our nIMParton16 is quite similar to EPS09 in the range of $10^{-4}<x_{B}<10^{-2}$, which happens to cover the range of the experimental data. The structures of rapidity distributions are similar, except for the CT10 model prediction. Owing to the considerable nuclei medium effects in Pb, the predictions of meson rapidity distributions by different models are suppressed a lot due to the nucleus shadowing effect of gluon distributions. In Fig.~\ref{Fig:4_3}, we show the corresponding predictions for the rapidity distributions of $J/\Psi$ photoproduction cross section in pp collisions at $\sqrt{s}~=~7~\rm{TeV}$. We present the predictions of different PDFs with the experimental data from the ALICE Collaboration~\cite{Abelev20131273, Abbas2013}. Since the rapidity interval corresponds to a wide range of $W_{\gamma p}$, the differences which appear in the predictions are amplified in the cross section rapidity distributions. We  find that our nIMParton16 modifications for $^{208}\rm{Pb}$ reproduce the PbPb UPCs reasonably like the EPS09 modifications. The nIMParton16 model is based on a combination of models proposed by the author in Ref.~\cite{Wang20171}. Based on the dynamical PDFs model and the model of nuclear medium effects, the nuclear modifications predict the nuclear correction of the deuteron, consistent with the measurement at JLab~\cite{PhysRevC.92.015211}.  

\section{Conclusion}
\label{SecV}
We conclude that diffractive $J/\Psi$ and $\Upsilon$ photoproductions at HERA, LHCb and ALICE offer us a unique opportunity to distinguish between the various gluon distributions in the kinematic region ($Q\sim M_{J/\Psi(\Upsilon)}/2$ and $x=M_{J/\Psi(\Upsilon)}^2/W_{\gamma p}^2$) where they are obviously distinct. The model of calculating the cross section of $\gamma p \to pV$ works well. The formulas are simply related to gluon distribution, which is a good approximation for testing gluon distribution models. Dynamical gluon distribution from IMParton at small $x$ is in agreement with exclusive $J/\Psi$ photoproduction in the LO approach, which means that the dynamical gluon distribution from nonlinear corrected DGLAP equations is reasonable in the small x region. Nuclear modifications of gluon distributions from nIMParton16 behave similarly to the EPS09 model in the small x region and it works well with large rapidity data for PbPb collisions data. The nIMParton16 is model-dependent and has only two free parameters. We hope more nuclear data at small x will improve the precision of nuclear medium corrections. In the future, there will be more experiments for exclusive meson production that covers smaller Bjorken-x. We look forward to more experiments on gluon detection in protons and nuclei in order to reveal more evidence of the gluon recombination effect.

\noindent{\bf Acknowledgments}:
The authors would like to thank professor Wei Zhu for helpful and fruitful suggestion. One of us (Qiang Fu) thanks Yaping Xie for help in checking of some calculations, and Rong Wang for the fruitful discussion on the parton distribution functions. This work is partly supported by the Key Research Program of Frontier Sciences, CAS, under the Grants Number NO. QYZDY-SSW-SLH006 of Chinese Academy of Sciences.

%


\begin{thebibliography}{43}%
\makeatletter
\providecommand \@ifxundefined [1]{%
 \@ifx{#1\undefined}
}%
\providecommand \@ifnum [1]{%
 \ifnum #1\expandafter \@firstoftwo
 \else \expandafter \@secondoftwo
 \fi
}%
\providecommand \@ifx [1]{%
 \ifx #1\expandafter \@firstoftwo
 \else \expandafter \@secondoftwo
 \fi
}%
\providecommand \natexlab [1]{#1}%
\providecommand \enquote  [1]{``#1''}%
\providecommand \bibnamefont  [1]{#1}%
\providecommand \bibfnamefont [1]{#1}%
\providecommand \citenamefont [1]{#1}%
\providecommand \href@noop [0]{\@secondoftwo}%
\providecommand \href [0]{\begingroup \@sanitize@url \@href}%
\providecommand \@href[1]{\@@startlink{#1}\@@href}%
\providecommand \@@href[1]{\endgroup#1\@@endlink}%
\providecommand \@sanitize@url [0]{\catcode `\\12\catcode `\$12\catcode
  `\&12\catcode `\#12\catcode `\^12\catcode `\_12\catcode `\%12\relax}%
\providecommand \@@startlink[1]{}%
\providecommand \@@endlink[0]{}%
\providecommand \url  [0]{\begingroup\@sanitize@url \@url }%
\providecommand \@url [1]{\endgroup\@href {#1}{\urlprefix }}%
\providecommand \urlprefix  [0]{URL }%
\providecommand \Eprint [0]{\href }%
\providecommand \doibase [0]{http://dx.doi.org/}%
\providecommand \selectlanguage [0]{\@gobble}%
\providecommand \bibinfo  [0]{\@secondoftwo}%
\providecommand \bibfield  [0]{\@secondoftwo}%
\providecommand \translation [1]{[#1]}%
\providecommand \BibitemOpen [0]{}%
\providecommand \bibitemStop [0]{}%
\providecommand \bibitemNoStop [0]{.\EOS\space}%
\providecommand \EOS [0]{\spacefactor3000\relax}%
\providecommand \BibitemShut  [1]{\csname bibitem#1\endcsname}%
\let\auto@bib@innerbib\@empty
\bibitem [{\citenamefont {Aaij}\ \emph {et~al.}(2014)\citenamefont {Aaij},
  \citenamefont {Adeva}, \citenamefont {Adinolfi},\ and\ \citenamefont
  {etc...}}]{0954-3899-41-5-055002}%
  \BibitemOpen
  \bibfield  {author} {\bibinfo {author} {\bibfnamefont {R.}~\bibnamefont
  {Aaij}}, \bibinfo {author} {\bibfnamefont {B.}~\bibnamefont {Adeva}},
  \bibinfo {author} {\bibfnamefont {M.}~\bibnamefont {Adinolfi}}, \ and\
  \bibinfo {author} {\bibnamefont {etc...}},\ }\href
  {http://stacks.iop.org/0954-3899/41/i=5/a=055002} {\bibfield  {journal}
  {\bibinfo  {journal} {Journal of Physics G: Nuclear and Particle Physics}\
  }\textbf {\bibinfo {volume} {41}},\ \bibinfo {pages} {055002} (\bibinfo
  {year} {2014})}\BibitemShut {NoStop}%
\bibitem [{\citenamefont {Aaij}\ \emph {et~al.}(2015)\citenamefont {Aaij},
  \citenamefont {Adeva}, \citenamefont {Adinolfi},\ and\ \citenamefont
  {{etal}}}]{Aaij2015}%
  \BibitemOpen
  \bibfield  {author} {\bibinfo {author} {\bibfnamefont {R.}~\bibnamefont
  {Aaij}}, \bibinfo {author} {\bibfnamefont {B.}~\bibnamefont {Adeva}},
  \bibinfo {author} {\bibfnamefont {M.}~\bibnamefont {Adinolfi}}, \ and\
  \bibinfo {author} {\bibnamefont {{etal}}},\ }\href {\doibase
  10.1007/JHEP09(2015)084} {\bibfield  {journal} {\bibinfo  {journal} {Journal
  of High Energy Physics}\ }\textbf {\bibinfo {volume} {2015}},\ \bibinfo
  {pages} {84} (\bibinfo {year} {2015})}\BibitemShut {NoStop}%
\bibitem [{\citenamefont {Chen}\ \emph {et~al.}(2014)\citenamefont {Chen},
  \citenamefont {Ruan}, \citenamefont {Wang}, \citenamefont {Zhang},\ and\
  \citenamefont {Zhu}}]{doi:10.1142/S0218301314500578}%
  \BibitemOpen
  \bibfield  {author} {\bibinfo {author} {\bibfnamefont {X.}~\bibnamefont
  {Chen}}, \bibinfo {author} {\bibfnamefont {J.}~\bibnamefont {Ruan}}, \bibinfo
  {author} {\bibfnamefont {R.}~\bibnamefont {Wang}}, \bibinfo {author}
  {\bibfnamefont {P.}~\bibnamefont {Zhang}}, \ and\ \bibinfo {author}
  {\bibfnamefont {W.}~\bibnamefont {Zhu}},\ }\href {\doibase
  10.1142/S0218301314500578} {\bibfield  {journal} {\bibinfo  {journal}
  {International Journal of Modern Physics E}\ }\textbf {\bibinfo {volume}
  {23}},\ \bibinfo {pages} {1450057} (\bibinfo {year} {2014})},\ \Eprint
  {http://arxiv.org/abs/http://www.worldscientific.com/doi/pdf/10.1142/S0218301314500578}
  {http://www.worldscientific.com/doi/pdf/10.1142/S0218301314500578}
  \BibitemShut {NoStop}%
\bibitem [{\citenamefont {Gon\ifmmode~\mbox{\c{c}}\else \c{c}\fi{}alves}\ and\
  \citenamefont {Bertulani}(2002)}]{PhysRevC.65.054905}%
  \BibitemOpen
  \bibfield  {author} {\bibinfo {author} {\bibfnamefont {V.~P.}\ \bibnamefont
  {Gon\ifmmode~\mbox{\c{c}}\else \c{c}\fi{}alves}}\ and\ \bibinfo {author}
  {\bibfnamefont {C.~A.}\ \bibnamefont {Bertulani}},\ }\href {\doibase
  10.1103/PhysRevC.65.054905} {\bibfield  {journal} {\bibinfo  {journal} {Phys.
  Rev. C}\ }\textbf {\bibinfo {volume} {65}},\ \bibinfo {pages} {054905}
  (\bibinfo {year} {2002})}\BibitemShut {NoStop}%
\bibitem [{\citenamefont {Fermi}(1925)}]{Fermi1925}%
  \BibitemOpen
  \bibfield  {author} {\bibinfo {author} {\bibfnamefont {E.}~\bibnamefont
  {Fermi}},\ }\href {\doibase 10.1007/BF02961914} {\bibfield  {journal}
  {\bibinfo  {journal} {Il Nuovo Cimento (1924-1942)}\ }\textbf {\bibinfo
  {volume} {2}},\ \bibinfo {pages} {143} (\bibinfo {year} {1925})}\BibitemShut
  {NoStop}%
\bibitem [{\citenamefont {Bertulani}\ and\ \citenamefont
  {Baur}(1988)}]{BERTULANI1988299}%
  \BibitemOpen
  \bibfield  {author} {\bibinfo {author} {\bibfnamefont {C.~A.}\ \bibnamefont
  {Bertulani}}\ and\ \bibinfo {author} {\bibfnamefont {G.}~\bibnamefont
  {Baur}},\ }\href {\doibase http://dx.doi.org/10.1016/0370-1573(88)90142-1}
  {\bibfield  {journal} {\bibinfo  {journal} {Physics Reports}\ }\textbf
  {\bibinfo {volume} {163}},\ \bibinfo {pages} {299 } (\bibinfo {year}
  {1988})}\BibitemShut {NoStop}%
\bibitem [{\citenamefont {Cahn}\ and\ \citenamefont
  {Jackson}(1990)}]{PhysRevD.42.3690}%
  \BibitemOpen
  \bibfield  {author} {\bibinfo {author} {\bibfnamefont {R.~N.}\ \bibnamefont
  {Cahn}}\ and\ \bibinfo {author} {\bibfnamefont {J.~D.}\ \bibnamefont
  {Jackson}},\ }\href {\doibase 10.1103/PhysRevD.42.3690} {\bibfield  {journal}
  {\bibinfo  {journal} {Phys. Rev. D}\ }\textbf {\bibinfo {volume} {42}},\
  \bibinfo {pages} {3690} (\bibinfo {year} {1990})}\BibitemShut {NoStop}%
\bibitem [{\citenamefont {Baur}\ and\ \citenamefont
  {Filho}(1990)}]{BAUR1990786}%
  \BibitemOpen
  \bibfield  {author} {\bibinfo {author} {\bibfnamefont {G.}~\bibnamefont
  {Baur}}\ and\ \bibinfo {author} {\bibfnamefont {L.}~\bibnamefont {Filho}},\
  }\href {\doibase http://dx.doi.org/10.1016/0375-9474(90)90191-N} {\bibfield
  {journal} {\bibinfo  {journal} {Nuclear Physics A}\ }\textbf {\bibinfo
  {volume} {518}},\ \bibinfo {pages} {786 } (\bibinfo {year}
  {1990})}\BibitemShut {NoStop}%
\bibitem [{\citenamefont {Ryskin}\ \emph {et~al.}(1997)\citenamefont {Ryskin},
  \citenamefont {Roberts}, \citenamefont {Martin},\ and\ \citenamefont
  {Levin}}]{Ryskin1997}%
  \BibitemOpen
  \bibfield  {author} {\bibinfo {author} {\bibfnamefont {M.~G.}\ \bibnamefont
  {Ryskin}}, \bibinfo {author} {\bibfnamefont {R.~G.}\ \bibnamefont {Roberts}},
  \bibinfo {author} {\bibfnamefont {A.~D.}\ \bibnamefont {Martin}}, \ and\
  \bibinfo {author} {\bibfnamefont {E.~M.}\ \bibnamefont {Levin}},\ }\href
  {\doibase 10.1007/s002880050547} {\bibfield  {journal} {\bibinfo  {journal}
  {Zeitschrift f{\"u}r Physik C Particles and Fields}\ }\textbf {\bibinfo
  {volume} {76}},\ \bibinfo {pages} {231} (\bibinfo {year} {1997})}\BibitemShut
  {NoStop}%
\bibitem [{\citenamefont {Adeluyi}\ and\ \citenamefont
  {Nguyen}(2013)}]{PhysRevC.87.027901}%
  \BibitemOpen
  \bibfield  {author} {\bibinfo {author} {\bibfnamefont {A.}~\bibnamefont
  {Adeluyi}}\ and\ \bibinfo {author} {\bibfnamefont {T.}~\bibnamefont
  {Nguyen}},\ }\href {\doibase 10.1103/PhysRevC.87.027901} {\bibfield
  {journal} {\bibinfo  {journal} {Phys. Rev. C}\ }\textbf {\bibinfo {volume}
  {87}},\ \bibinfo {pages} {027901} (\bibinfo {year} {2013})}\BibitemShut
  {NoStop}%
\bibitem [{\citenamefont {Klein}\ and\ \citenamefont
  {Nystrand}(2004)}]{PhysRevLett.92.142003}%
  \BibitemOpen
  \bibfield  {author} {\bibinfo {author} {\bibfnamefont {S.~R.}\ \bibnamefont
  {Klein}}\ and\ \bibinfo {author} {\bibfnamefont {J.}~\bibnamefont
  {Nystrand}},\ }\href {\doibase 10.1103/PhysRevLett.92.142003} {\bibfield
  {journal} {\bibinfo  {journal} {Phys. Rev. Lett.}\ }\textbf {\bibinfo
  {volume} {92}},\ \bibinfo {pages} {142003} (\bibinfo {year}
  {2004})}\BibitemShut {NoStop}%
\bibitem [{\citenamefont {Adeluyi}\ and\ \citenamefont
  {Bertulani}(2011)}]{PhysRevC.84.024916}%
  \BibitemOpen
  \bibfield  {author} {\bibinfo {author} {\bibfnamefont {A.}~\bibnamefont
  {Adeluyi}}\ and\ \bibinfo {author} {\bibfnamefont {C.~A.}\ \bibnamefont
  {Bertulani}},\ }\href {\doibase 10.1103/PhysRevC.84.024916} {\bibfield
  {journal} {\bibinfo  {journal} {Phys. Rev. C}\ }\textbf {\bibinfo {volume}
  {84}},\ \bibinfo {pages} {024916} (\bibinfo {year} {2011})}\BibitemShut
  {NoStop}%
\bibitem [{\citenamefont {Gon{\c{c}}alves}\ \emph {et~al.}(2016)\citenamefont
  {Gon{\c{c}}alves}, \citenamefont {Martins},\ and\ \citenamefont
  {Sauter}}]{Gonçalves2016}%
  \BibitemOpen
  \bibfield  {author} {\bibinfo {author} {\bibfnamefont {V.~P.}\ \bibnamefont
  {Gon{\c{c}}alves}}, \bibinfo {author} {\bibfnamefont {L.~A.~S.}\ \bibnamefont
  {Martins}}, \ and\ \bibinfo {author} {\bibfnamefont {W.~K.}\ \bibnamefont
  {Sauter}},\ }\href {\doibase 10.1140/epjc/s10052-016-3917-z} {\bibfield
  {journal} {\bibinfo  {journal} {The European Physical Journal C}\ }\textbf
  {\bibinfo {volume} {76}},\ \bibinfo {pages} {97} (\bibinfo {year}
  {2016})}\BibitemShut {NoStop}%
\bibitem [{\citenamefont {Zhu}(1999)}]{WEIZHU1999245}%
  \BibitemOpen
  \bibfield  {author} {\bibinfo {author} {\bibfnamefont {W.}~\bibnamefont
  {Zhu}},\ }\href {\doibase http://dx.doi.org/10.1016/S0550-3213(99)00237-0}
  {\bibfield  {journal} {\bibinfo  {journal} {Nuclear Physics B}\ }\textbf
  {\bibinfo {volume} {551}},\ \bibinfo {pages} {245 } (\bibinfo {year}
  {1999})}\BibitemShut {NoStop}%
\bibitem [{\citenamefont {Zhu}\ and\ \citenamefont {Ruan}(1999)}]{ZHU1999378}%
  \BibitemOpen
  \bibfield  {author} {\bibinfo {author} {\bibfnamefont {W.}~\bibnamefont
  {Zhu}}\ and\ \bibinfo {author} {\bibfnamefont {J.}~\bibnamefont {Ruan}},\
  }\href {\doibase http://dx.doi.org/10.1016/S0550-3213(99)00461-7} {\bibfield
  {journal} {\bibinfo  {journal} {Nuclear Physics B}\ }\textbf {\bibinfo
  {volume} {559}},\ \bibinfo {pages} {378 } (\bibinfo {year}
  {1999})}\BibitemShut {NoStop}%
\bibitem [{\citenamefont {Zhu}\ and\ \citenamefont {Shen}(2005)}]{Zhu:2004xj}%
  \BibitemOpen
  \bibfield  {author} {\bibinfo {author} {\bibfnamefont {W.}~\bibnamefont
  {Zhu}}\ and\ \bibinfo {author} {\bibfnamefont {Z.-q.}\ \bibnamefont {Shen}},\
  }\href@noop {} {\bibfield  {journal} {\bibinfo  {journal} {HEP. amp; NP.
  Vol.}\ }\textbf {\bibinfo {volume} {2}},\ \bibinfo {pages} {109} (\bibinfo
  {year} {2005})},\ \Eprint {http://arxiv.org/abs/hep-ph/0406213}
  {arXiv:hep-ph/0406213 [hep-ph]} \BibitemShut {NoStop}%
\bibitem [{\citenamefont {Gelis}\ \emph {et~al.}(2010)\citenamefont {Gelis},
  \citenamefont {Iancu}, \citenamefont {Jalilian-Marian},\ and\ \citenamefont
  {Venugopalan}}]{doi:10.1146/annurev.nucl.010909.083629}%
  \BibitemOpen
  \bibfield  {author} {\bibinfo {author} {\bibfnamefont {F.}~\bibnamefont
  {Gelis}}, \bibinfo {author} {\bibfnamefont {E.}~\bibnamefont {Iancu}},
  \bibinfo {author} {\bibfnamefont {J.}~\bibnamefont {Jalilian-Marian}}, \ and\
  \bibinfo {author} {\bibfnamefont {R.}~\bibnamefont {Venugopalan}},\ }\href
  {\doibase 10.1146/annurev.nucl.010909.083629} {\bibfield  {journal} {\bibinfo
   {journal} {Annual Review of Nuclear and Particle Science}\ }\textbf
  {\bibinfo {volume} {60}},\ \bibinfo {pages} {463} (\bibinfo {year} {2010})},\
  \Eprint
  {http://arxiv.org/abs/https://doi.org/10.1146/annurev.nucl.010909.083629}
  {https://doi.org/10.1146/annurev.nucl.010909.083629} \BibitemShut {NoStop}%
\bibitem [{\citenamefont {Wang}\ and\ \citenamefont
  {Chen}(2017)}]{1674-1137-41-5-053103}%
  \BibitemOpen
  \bibfield  {author} {\bibinfo {author} {\bibfnamefont {R.}~\bibnamefont
  {Wang}}\ and\ \bibinfo {author} {\bibfnamefont {X.-R.}\ \bibnamefont
  {Chen}},\ }\href {http://stacks.iop.org/1674-1137/41/i=5/a=053103} {\bibfield
   {journal} {\bibinfo  {journal} {Chinese Physics C}\ }\textbf {\bibinfo
  {volume} {41}},\ \bibinfo {pages} {053103} (\bibinfo {year}
  {2017})}\BibitemShut {NoStop}%
\bibitem [{\citenamefont {Harland-Lang}\ \emph {et~al.}(2015)\citenamefont
  {Harland-Lang}, \citenamefont {Martin}, \citenamefont {Motylinski},\ and\
  \citenamefont {Thorne}}]{Harland-Lang2015}%
  \BibitemOpen
  \bibfield  {author} {\bibinfo {author} {\bibfnamefont {L.~A.}\ \bibnamefont
  {Harland-Lang}}, \bibinfo {author} {\bibfnamefont {A.~D.}\ \bibnamefont
  {Martin}}, \bibinfo {author} {\bibfnamefont {P.}~\bibnamefont {Motylinski}},
  \ and\ \bibinfo {author} {\bibfnamefont {R.~S.}\ \bibnamefont {Thorne}},\
  }\href {\doibase 10.1140/epjc/s10052-015-3397-6} {\bibfield  {journal}
  {\bibinfo  {journal} {The European Physical Journal C}\ }\textbf {\bibinfo
  {volume} {75}},\ \bibinfo {pages} {204} (\bibinfo {year} {2015})}\BibitemShut
  {NoStop}%
\bibitem [{\citenamefont {Gl{\"u}ck}\ \emph {et~al.}(2008)\citenamefont
  {Gl{\"u}ck}, \citenamefont {Jimenez-Delgado},\ and\ \citenamefont
  {Reya}}]{Gluck2008}%
  \BibitemOpen
  \bibfield  {author} {\bibinfo {author} {\bibfnamefont {M.}~\bibnamefont
  {Gl{\"u}ck}}, \bibinfo {author} {\bibfnamefont {P.}~\bibnamefont
  {Jimenez-Delgado}}, \ and\ \bibinfo {author} {\bibfnamefont {E.}~\bibnamefont
  {Reya}},\ }\href {\doibase 10.1140/epjc/s10052-007-0462-9} {\bibfield
  {journal} {\bibinfo  {journal} {The European Physical Journal C}\ }\textbf
  {\bibinfo {volume} {53}},\ \bibinfo {pages} {355} (\bibinfo {year}
  {2008})}\BibitemShut {NoStop}%
\bibitem [{\citenamefont {Lai}\ \emph {et~al.}(2010)\citenamefont {Lai},
  \citenamefont {Guzzi}, \citenamefont {Huston}, \citenamefont {Li},
  \citenamefont {Nadolsky}, \citenamefont {Pumplin},\ and\ \citenamefont
  {Yuan}}]{PhysRevD.82.074024}%
  \BibitemOpen
  \bibfield  {author} {\bibinfo {author} {\bibfnamefont {H.-L.}\ \bibnamefont
  {Lai}}, \bibinfo {author} {\bibfnamefont {M.}~\bibnamefont {Guzzi}}, \bibinfo
  {author} {\bibfnamefont {J.}~\bibnamefont {Huston}}, \bibinfo {author}
  {\bibfnamefont {Z.}~\bibnamefont {Li}}, \bibinfo {author} {\bibfnamefont
  {P.~M.}\ \bibnamefont {Nadolsky}}, \bibinfo {author} {\bibfnamefont
  {J.}~\bibnamefont {Pumplin}}, \ and\ \bibinfo {author} {\bibfnamefont
  {C.-P.}\ \bibnamefont {Yuan}},\ }\href {\doibase 10.1103/PhysRevD.82.074024}
  {\bibfield  {journal} {\bibinfo  {journal} {Phys. Rev. D}\ }\textbf {\bibinfo
  {volume} {82}},\ \bibinfo {pages} {074024} (\bibinfo {year}
  {2010})}\BibitemShut {NoStop}%
\bibitem [{\citenamefont {Brodsky}\ \emph {et~al.}(1994)\citenamefont
  {Brodsky}, \citenamefont {Frankfurt}, \citenamefont {Gunion}, \citenamefont
  {Mueller},\ and\ \citenamefont {Strikman}}]{PhysRevD.50.3134}%
  \BibitemOpen
  \bibfield  {author} {\bibinfo {author} {\bibfnamefont {S.~J.}\ \bibnamefont
  {Brodsky}}, \bibinfo {author} {\bibfnamefont {L.}~\bibnamefont {Frankfurt}},
  \bibinfo {author} {\bibfnamefont {J.~F.}\ \bibnamefont {Gunion}}, \bibinfo
  {author} {\bibfnamefont {A.~H.}\ \bibnamefont {Mueller}}, \ and\ \bibinfo
  {author} {\bibfnamefont {M.}~\bibnamefont {Strikman}},\ }\href {\doibase
  10.1103/PhysRevD.50.3134} {\bibfield  {journal} {\bibinfo  {journal} {Phys.
  Rev. D}\ }\textbf {\bibinfo {volume} {50}},\ \bibinfo {pages} {3134}
  (\bibinfo {year} {1994})}\BibitemShut {NoStop}%
\bibitem [{\citenamefont {Frankfurt}\ \emph {et~al.}(1998)\citenamefont
  {Frankfurt}, \citenamefont {Koepf},\ and\ \citenamefont
  {Strikman}}]{PhysRevD.57.512}%
  \BibitemOpen
  \bibfield  {author} {\bibinfo {author} {\bibfnamefont {L.}~\bibnamefont
  {Frankfurt}}, \bibinfo {author} {\bibfnamefont {W.}~\bibnamefont {Koepf}}, \
  and\ \bibinfo {author} {\bibfnamefont {M.}~\bibnamefont {Strikman}},\ }\href
  {\doibase 10.1103/PhysRevD.57.512} {\bibfield  {journal} {\bibinfo  {journal}
  {Phys. Rev. D}\ }\textbf {\bibinfo {volume} {57}},\ \bibinfo {pages} {512}
  (\bibinfo {year} {1998})}\BibitemShut {NoStop}%
\bibitem [{\citenamefont {Eskola}\ \emph {et~al.}(2009)\citenamefont {Eskola},
  \citenamefont {Paukkunen},\ and\ \citenamefont
  {Salgado}}]{1126-6708-2009-04-065}%
  \BibitemOpen
  \bibfield  {author} {\bibinfo {author} {\bibfnamefont {K.}~\bibnamefont
  {Eskola}}, \bibinfo {author} {\bibfnamefont {H.}~\bibnamefont {Paukkunen}}, \
  and\ \bibinfo {author} {\bibfnamefont {C.}~\bibnamefont {Salgado}},\ }\href
  {http://stacks.iop.org/1126-6708/2009/i=04/a=065} {\bibfield  {journal}
  {\bibinfo  {journal} {Journal of High Energy Physics}\ }\textbf {\bibinfo
  {volume} {2009}},\ \bibinfo {pages} {065} (\bibinfo {year}
  {2009})}\BibitemShut {NoStop}%
\bibitem [{\citenamefont {Wang}\ \emph {et~al.}(2017)\citenamefont {Wang},
  \citenamefont {Chen},\ and\ \citenamefont {Fu}}]{Wang20171}%
  \BibitemOpen
  \bibfield  {author} {\bibinfo {author} {\bibfnamefont {R.}~\bibnamefont
  {Wang}}, \bibinfo {author} {\bibfnamefont {X.}~\bibnamefont {Chen}}, \ and\
  \bibinfo {author} {\bibfnamefont {Q.}~\bibnamefont {Fu}},\ }\href {\doibase
  https://doi.org/10.1016/j.nuclphysb.2017.04.008} {\bibfield  {journal}
  {\bibinfo  {journal} {Nuclear Physics B}\ }\textbf {\bibinfo {volume}
  {920}},\ \bibinfo {pages} {1 } (\bibinfo {year} {2017})}\BibitemShut
  {NoStop}%
\bibitem [{\citenamefont {Alexa}\ \emph {et~al.}(2013)\citenamefont {Alexa},
  \citenamefont {Andreev}, \citenamefont {Baghdasaryan},\ and\ \citenamefont
  {etc...}}]{Alexa2013}%
  \BibitemOpen
  \bibfield  {author} {\bibinfo {author} {\bibfnamefont {C.}~\bibnamefont
  {Alexa}}, \bibinfo {author} {\bibfnamefont {V.}~\bibnamefont {Andreev}},
  \bibinfo {author} {\bibfnamefont {A.}~\bibnamefont {Baghdasaryan}}, \ and\
  \bibinfo {author} {\bibnamefont {etc...}},\ }\href {\doibase
  10.1140/epjc/s10052-013-2466-y} {\bibfield  {journal} {\bibinfo  {journal}
  {The European Physical Journal C}\ }\textbf {\bibinfo {volume} {73}},\
  \bibinfo {pages} {2466} (\bibinfo {year} {2013})}\BibitemShut {NoStop}%
\bibitem [{\citenamefont {Chekanov~et al.}(2002)}]{Chekanovetal.2002}%
  \BibitemOpen
  \bibfield  {author} {\bibinfo {author} {\bibfnamefont {S.}~\bibnamefont
  {Chekanov~et al.}},\ }\href {\doibase 10.1007/s10052-002-0953-7} {\bibfield
  {journal} {\bibinfo  {journal} {The European Physical Journal C - Particles
  and Fields}\ }\textbf {\bibinfo {volume} {24}},\ \bibinfo {pages} {345}
  (\bibinfo {year} {2002})}\BibitemShut {NoStop}%
\bibitem [{\citenamefont {Chekanov}\ \emph {et~al.}(2004)\citenamefont
  {Chekanov}, \citenamefont {Derrick}, \citenamefont {Loizides},\ and\
  \citenamefont {etc...}}]{Chekanov20043}%
  \BibitemOpen
  \bibfield  {author} {\bibinfo {author} {\bibfnamefont {S.}~\bibnamefont
  {Chekanov}}, \bibinfo {author} {\bibfnamefont {M.}~\bibnamefont {Derrick}},
  \bibinfo {author} {\bibfnamefont {J.}~\bibnamefont {Loizides}}, \ and\
  \bibinfo {author} {\bibnamefont {etc...}},\ }\href {\doibase
  https://doi.org/10.1016/j.nuclphysb.2004.06.034} {\bibfield  {journal}
  {\bibinfo  {journal} {Nuclear Physics B}\ }\textbf {\bibinfo {volume}
  {695}},\ \bibinfo {pages} {3 } (\bibinfo {year} {2004})}\BibitemShut
  {NoStop}%
\bibitem [{\citenamefont {Abelev}\ \emph {et~al.}(2014)\citenamefont {Abelev},
  \citenamefont {Adam},\ and\ \citenamefont
  {Adamov\'a}}]{PhysRevLett.113.232504}%
  \BibitemOpen
  \bibfield  {author} {\bibinfo {author} {\bibfnamefont {B.}~\bibnamefont
  {Abelev}}, \bibinfo {author} {\bibfnamefont {J.}~\bibnamefont {Adam}}, \ and\
  \bibinfo {author} {\bibfnamefont {D.~e.}\ \bibnamefont {Adamov\'a}} (\bibinfo
  {collaboration} {ALICE Collaboration}),\ }\href {\doibase
  10.1103/PhysRevLett.113.232504} {\bibfield  {journal} {\bibinfo  {journal}
  {Phys. Rev. Lett.}\ }\textbf {\bibinfo {volume} {113}},\ \bibinfo {pages}
  {232504} (\bibinfo {year} {2014})}\BibitemShut {NoStop}%
\bibitem [{\citenamefont {Adloff}\ \emph {et~al.}(2000)\citenamefont {Adloff},
  \citenamefont {Andreev}, \citenamefont {Andrieu},\ and\ \citenamefont
  {etc...}}]{Adloff200023}%
  \BibitemOpen
  \bibfield  {author} {\bibinfo {author} {\bibfnamefont {C.}~\bibnamefont
  {Adloff}}, \bibinfo {author} {\bibfnamefont {V.}~\bibnamefont {Andreev}},
  \bibinfo {author} {\bibfnamefont {B.}~\bibnamefont {Andrieu}}, \ and\
  \bibinfo {author} {\bibnamefont {etc...}},\ }\href {\doibase
  https://doi.org/10.1016/S0370-2693(00)00530-X} {\bibfield  {journal}
  {\bibinfo  {journal} {Physics Letters B}\ }\textbf {\bibinfo {volume}
  {483}},\ \bibinfo {pages} {23 } (\bibinfo {year} {2000})}\BibitemShut
  {NoStop}%
\bibitem [{\citenamefont {Aaij}\ \emph {et~al.}(2013)\citenamefont {Aaij},
  \citenamefont {Beteta},\ and\ \citenamefont
  {etc...}}]{0954-3899-40-4-045001}%
  \BibitemOpen
  \bibfield  {author} {\bibinfo {author} {\bibfnamefont {R.}~\bibnamefont
  {Aaij}}, \bibinfo {author} {\bibfnamefont {C.~A.}\ \bibnamefont {Beteta}}, \
  and\ \bibinfo {author} {\bibnamefont {etc...}},\ }\href
  {http://stacks.iop.org/0954-3899/40/i=4/a=045001} {\bibfield  {journal}
  {\bibinfo  {journal} {Journal of Physics G: Nuclear and Particle Physics}\
  }\textbf {\bibinfo {volume} {40}},\ \bibinfo {pages} {045001} (\bibinfo
  {year} {2013})}\BibitemShut {NoStop}%
\bibitem [{\citenamefont {Breitweg}\ \emph {et~al.}(1998)\citenamefont
  {Breitweg}, \citenamefont {Derrick}, \citenamefont {Krakauer},\ and\
  \citenamefont {etc...}}]{Breitweg1998432}%
  \BibitemOpen
  \bibfield  {author} {\bibinfo {author} {\bibfnamefont {J.}~\bibnamefont
  {Breitweg}}, \bibinfo {author} {\bibfnamefont {M.}~\bibnamefont {Derrick}},
  \bibinfo {author} {\bibfnamefont {D.}~\bibnamefont {Krakauer}}, \ and\
  \bibinfo {author} {\bibnamefont {etc...}},\ }\href {\doibase
  https://doi.org/10.1016/S0370-2693(98)01081-8} {\bibfield  {journal}
  {\bibinfo  {journal} {Physics Letters B}\ }\textbf {\bibinfo {volume}
  {437}},\ \bibinfo {pages} {432 } (\bibinfo {year} {1998})}\BibitemShut
  {NoStop}%
\bibitem [{\citenamefont {Chekanov}\ \emph {et~al.}(2009)\citenamefont
  {Chekanov}, \citenamefont {Derrick}, \citenamefont {Magill},\ and\
  \citenamefont {etc...}}]{Chekanov20094}%
  \BibitemOpen
  \bibfield  {author} {\bibinfo {author} {\bibfnamefont {S.}~\bibnamefont
  {Chekanov}}, \bibinfo {author} {\bibfnamefont {M.}~\bibnamefont {Derrick}},
  \bibinfo {author} {\bibfnamefont {S.}~\bibnamefont {Magill}}, \ and\ \bibinfo
  {author} {\bibnamefont {etc...}},\ }\href {\doibase
  https://doi.org/10.1016/j.physletb.2009.07.066} {\bibfield  {journal}
  {\bibinfo  {journal} {Physics Letters B}\ }\textbf {\bibinfo {volume}
  {680}},\ \bibinfo {pages} {4 } (\bibinfo {year} {2009})}\BibitemShut
  {NoStop}%
\bibitem [{\citenamefont {Vries}\ \emph {et~al.}(1987)\citenamefont {Vries},
  \citenamefont {Jager},\ and\ \citenamefont {Vries}}]{DEVRIES1987495}%
  \BibitemOpen
  \bibfield  {author} {\bibinfo {author} {\bibfnamefont {H.~D.}\ \bibnamefont
  {Vries}}, \bibinfo {author} {\bibfnamefont {C.~D.}\ \bibnamefont {Jager}}, \
  and\ \bibinfo {author} {\bibfnamefont {C.~D.}\ \bibnamefont {Vries}},\ }\href
  {\doibase http://dx.doi.org/10.1016/0092-640X(87)90013-1} {\bibfield
  {journal} {\bibinfo  {journal} {Atomic Data and Nuclear Data Tables}\
  }\textbf {\bibinfo {volume} {36}},\ \bibinfo {pages} {495 } (\bibinfo {year}
  {1987})}\BibitemShut {NoStop}%
\bibitem [{\citenamefont {Bertulani}\ and\ \citenamefont
  {Dolci}(2000)}]{BERTULANI2000527}%
  \BibitemOpen
  \bibfield  {author} {\bibinfo {author} {\bibfnamefont {C.}~\bibnamefont
  {Bertulani}}\ and\ \bibinfo {author} {\bibfnamefont {D.}~\bibnamefont
  {Dolci}},\ }\href {\doibase http://dx.doi.org/10.1016/S0375-9474(00)00172-X}
  {\bibfield  {journal} {\bibinfo  {journal} {Nuclear Physics A}\ }\textbf
  {\bibinfo {volume} {674}},\ \bibinfo {pages} {527 } (\bibinfo {year}
  {2000})}\BibitemShut {NoStop}%
\bibitem [{\citenamefont {Davies}\ and\ \citenamefont
  {Nix}(1976)}]{PhysRevC.14.1977}%
  \BibitemOpen
  \bibfield  {author} {\bibinfo {author} {\bibfnamefont {K.~T.~R.}\
  \bibnamefont {Davies}}\ and\ \bibinfo {author} {\bibfnamefont {J.~R.}\
  \bibnamefont {Nix}},\ }\href {\doibase 10.1103/PhysRevC.14.1977} {\bibfield
  {journal} {\bibinfo  {journal} {Phys. Rev. C}\ }\textbf {\bibinfo {volume}
  {14}},\ \bibinfo {pages} {1977} (\bibinfo {year} {1976})}\BibitemShut
  {NoStop}%
\bibitem [{\citenamefont {Thomas}\ \emph {et~al.}(2016)\citenamefont {Thomas},
  \citenamefont {Bertulani}, \citenamefont {Brady}, \citenamefont {Clark},
  \citenamefont {Godat},\ and\ \citenamefont {Olness}}]{Thomas:2016oms}%
  \BibitemOpen
  \bibfield  {author} {\bibinfo {author} {\bibfnamefont {J.}~\bibnamefont
  {Thomas}}, \bibinfo {author} {\bibfnamefont {C.~A.}\ \bibnamefont
  {Bertulani}}, \bibinfo {author} {\bibfnamefont {N.}~\bibnamefont {Brady}},
  \bibinfo {author} {\bibfnamefont {D.~B.}\ \bibnamefont {Clark}}, \bibinfo
  {author} {\bibfnamefont {E.}~\bibnamefont {Godat}}, \ and\ \bibinfo {author}
  {\bibfnamefont {F.}~\bibnamefont {Olness}},\ }\href@noop {} {\  (\bibinfo
  {year} {2016})},\ \Eprint {http://arxiv.org/abs/1603.01919} {arXiv:1603.01919
  [hep-ph]} \BibitemShut {NoStop}%
\bibitem [{\citenamefont {Aubert}\ \emph {et~al.}(1983)\citenamefont {Aubert},
  \citenamefont {Bassompierre}, \citenamefont {Becks}, \citenamefont {Best},
  \citenamefont {Böhm}, \citenamefont {de~Bouard}, \citenamefont {Brasse},
  \citenamefont {Broll}, \citenamefont {Brown}, \citenamefont {Carr},
  \citenamefont {Clifft}, \citenamefont {Cobb}, \citenamefont {Coignet},
  \citenamefont {Combley}, \citenamefont {Court}, \citenamefont {D'Agostini},
  \citenamefont {Dau}, \citenamefont {Davies}, \citenamefont {Déclais},
  \citenamefont {Dobinson}, \citenamefont {Dosselli}, \citenamefont {Drees},
  \citenamefont {Edwards}, \citenamefont {Edwards}, \citenamefont {Favier},
  \citenamefont {Ferrero}, \citenamefont {Flauger}, \citenamefont {Gabathuler},
  \citenamefont {Gamet}, \citenamefont {Gayler}, \citenamefont {Gerhardt},
  \citenamefont {Gössling}, \citenamefont {Haas}, \citenamefont {Hamacher},
  \citenamefont {Hayman}, \citenamefont {Henckes}, \citenamefont {Korbel},
  \citenamefont {Landgraf}, \citenamefont {Leenen}, \citenamefont {Maire},
  \citenamefont {Minssieux}, \citenamefont {Mohr}, \citenamefont {Montgomery},
  \citenamefont {Moser}, \citenamefont {Mount}, \citenamefont {Norton},
  \citenamefont {McNicholas}, \citenamefont {Osborne}, \citenamefont {Payre},
  \citenamefont {Peroni}, \citenamefont {Pessard}, \citenamefont {Pietrzyk},
  \citenamefont {Rith}, \citenamefont {Schneegans}, \citenamefont {Sloan},
  \citenamefont {Stier}, \citenamefont {Stockhausen}, \citenamefont {Thénard},
  \citenamefont {Thompson}, \citenamefont {Urban}, \citenamefont {Villers},
  \citenamefont {Wahlen}, \citenamefont {Whalley}, \citenamefont {Williams},
  \citenamefont {Williams}, \citenamefont {Williamson},\ and\ \citenamefont
  {Wimpenny}}]{AUBERT1983275}%
  \BibitemOpen
  \bibfield  {author} {\bibinfo {author} {\bibfnamefont {J.}~\bibnamefont
  {Aubert}}, \bibinfo {author} {\bibfnamefont {G.}~\bibnamefont
  {Bassompierre}}, \bibinfo {author} {\bibfnamefont {K.}~\bibnamefont {Becks}},
  \bibinfo {author} {\bibfnamefont {C.}~\bibnamefont {Best}}, \bibinfo {author}
  {\bibfnamefont {E.}~\bibnamefont {Böhm}}, \bibinfo {author} {\bibfnamefont
  {X.}~\bibnamefont {de~Bouard}}, \bibinfo {author} {\bibfnamefont
  {F.}~\bibnamefont {Brasse}}, \bibinfo {author} {\bibfnamefont
  {C.}~\bibnamefont {Broll}}, \bibinfo {author} {\bibfnamefont
  {S.}~\bibnamefont {Brown}}, \bibinfo {author} {\bibfnamefont
  {J.}~\bibnamefont {Carr}}, \bibinfo {author} {\bibfnamefont {R.}~\bibnamefont
  {Clifft}}, \bibinfo {author} {\bibfnamefont {J.}~\bibnamefont {Cobb}},
  \bibinfo {author} {\bibfnamefont {G.}~\bibnamefont {Coignet}}, \bibinfo
  {author} {\bibfnamefont {F.}~\bibnamefont {Combley}}, \bibinfo {author}
  {\bibfnamefont {G.}~\bibnamefont {Court}}, \bibinfo {author} {\bibfnamefont
  {G.}~\bibnamefont {D'Agostini}}, \bibinfo {author} {\bibfnamefont
  {W.}~\bibnamefont {Dau}}, \bibinfo {author} {\bibfnamefont {J.}~\bibnamefont
  {Davies}}, \bibinfo {author} {\bibfnamefont {Y.}~\bibnamefont {Déclais}},
  \bibinfo {author} {\bibfnamefont {R.}~\bibnamefont {Dobinson}}, \bibinfo
  {author} {\bibfnamefont {U.}~\bibnamefont {Dosselli}}, \bibinfo {author}
  {\bibfnamefont {J.}~\bibnamefont {Drees}}, \bibinfo {author} {\bibfnamefont
  {A.}~\bibnamefont {Edwards}}, \bibinfo {author} {\bibfnamefont
  {M.}~\bibnamefont {Edwards}}, \bibinfo {author} {\bibfnamefont
  {J.}~\bibnamefont {Favier}}, \bibinfo {author} {\bibfnamefont
  {M.}~\bibnamefont {Ferrero}}, \bibinfo {author} {\bibfnamefont
  {W.}~\bibnamefont {Flauger}}, \bibinfo {author} {\bibfnamefont
  {E.}~\bibnamefont {Gabathuler}}, \bibinfo {author} {\bibfnamefont
  {R.}~\bibnamefont {Gamet}}, \bibinfo {author} {\bibfnamefont
  {J.}~\bibnamefont {Gayler}}, \bibinfo {author} {\bibfnamefont
  {V.}~\bibnamefont {Gerhardt}}, \bibinfo {author} {\bibfnamefont
  {C.}~\bibnamefont {Gössling}}, \bibinfo {author} {\bibfnamefont
  {J.}~\bibnamefont {Haas}}, \bibinfo {author} {\bibfnamefont {K.}~\bibnamefont
  {Hamacher}}, \bibinfo {author} {\bibfnamefont {P.}~\bibnamefont {Hayman}},
  \bibinfo {author} {\bibfnamefont {M.}~\bibnamefont {Henckes}}, \bibinfo
  {author} {\bibfnamefont {V.}~\bibnamefont {Korbel}}, \bibinfo {author}
  {\bibfnamefont {U.}~\bibnamefont {Landgraf}}, \bibinfo {author}
  {\bibfnamefont {M.}~\bibnamefont {Leenen}}, \bibinfo {author} {\bibfnamefont
  {M.}~\bibnamefont {Maire}}, \bibinfo {author} {\bibfnamefont
  {H.}~\bibnamefont {Minssieux}}, \bibinfo {author} {\bibfnamefont
  {W.}~\bibnamefont {Mohr}}, \bibinfo {author} {\bibfnamefont {H.}~\bibnamefont
  {Montgomery}}, \bibinfo {author} {\bibfnamefont {K.}~\bibnamefont {Moser}},
  \bibinfo {author} {\bibfnamefont {R.}~\bibnamefont {Mount}}, \bibinfo
  {author} {\bibfnamefont {P.}~\bibnamefont {Norton}}, \bibinfo {author}
  {\bibfnamefont {J.}~\bibnamefont {McNicholas}}, \bibinfo {author}
  {\bibfnamefont {A.}~\bibnamefont {Osborne}}, \bibinfo {author} {\bibfnamefont
  {P.}~\bibnamefont {Payre}}, \bibinfo {author} {\bibfnamefont
  {C.}~\bibnamefont {Peroni}}, \bibinfo {author} {\bibfnamefont
  {H.}~\bibnamefont {Pessard}}, \bibinfo {author} {\bibfnamefont
  {U.}~\bibnamefont {Pietrzyk}}, \bibinfo {author} {\bibfnamefont
  {K.}~\bibnamefont {Rith}}, \bibinfo {author} {\bibfnamefont {M.}~\bibnamefont
  {Schneegans}}, \bibinfo {author} {\bibfnamefont {T.}~\bibnamefont {Sloan}},
  \bibinfo {author} {\bibfnamefont {H.}~\bibnamefont {Stier}}, \bibinfo
  {author} {\bibfnamefont {W.}~\bibnamefont {Stockhausen}}, \bibinfo {author}
  {\bibfnamefont {J.}~\bibnamefont {Thénard}}, \bibinfo {author}
  {\bibfnamefont {J.}~\bibnamefont {Thompson}}, \bibinfo {author}
  {\bibfnamefont {L.}~\bibnamefont {Urban}}, \bibinfo {author} {\bibfnamefont
  {M.}~\bibnamefont {Villers}}, \bibinfo {author} {\bibfnamefont
  {H.}~\bibnamefont {Wahlen}}, \bibinfo {author} {\bibfnamefont
  {M.}~\bibnamefont {Whalley}}, \bibinfo {author} {\bibfnamefont
  {D.}~\bibnamefont {Williams}}, \bibinfo {author} {\bibfnamefont
  {W.}~\bibnamefont {Williams}}, \bibinfo {author} {\bibfnamefont
  {J.}~\bibnamefont {Williamson}}, \ and\ \bibinfo {author} {\bibfnamefont
  {S.}~\bibnamefont {Wimpenny}},\ }\href {\doibase
  http://dx.doi.org/10.1016/0370-2693(83)90437-9} {\bibfield  {journal}
  {\bibinfo  {journal} {Physics Letters B}\ }\textbf {\bibinfo {volume}
  {123}},\ \bibinfo {pages} {275 } (\bibinfo {year} {1983})}\BibitemShut
  {NoStop}%
\bibitem [{\citenamefont {Abelev}\ \emph {et~al.}(2013)\citenamefont {Abelev},
  \citenamefont {Adam}, \citenamefont {Adamová},\ and\ \citenamefont
  {etc...}}]{Abelev20131273}%
  \BibitemOpen
  \bibfield  {author} {\bibinfo {author} {\bibfnamefont {B.}~\bibnamefont
  {Abelev}}, \bibinfo {author} {\bibfnamefont {J.}~\bibnamefont {Adam}},
  \bibinfo {author} {\bibfnamefont {D.}~\bibnamefont {Adamová}}, \ and\
  \bibinfo {author} {\bibnamefont {etc...}},\ }\href {\doibase
  https://doi.org/10.1016/j.physletb.2012.11.059} {\bibfield  {journal}
  {\bibinfo  {journal} {Physics Letters B}\ }\textbf {\bibinfo {volume}
  {718}},\ \bibinfo {pages} {1273 } (\bibinfo {year} {2013})}\BibitemShut
  {NoStop}%
\bibitem [{\citenamefont {Abbas}\ \emph {et~al.}(2013)\citenamefont {Abbas},
  \citenamefont {Abelev}, \citenamefont {Adam},\ and\ \citenamefont
  {{etal}}}]{Abbas2013}%
  \BibitemOpen
  \bibfield  {author} {\bibinfo {author} {\bibfnamefont {E.}~\bibnamefont
  {Abbas}}, \bibinfo {author} {\bibfnamefont {B.}~\bibnamefont {Abelev}},
  \bibinfo {author} {\bibfnamefont {J.}~\bibnamefont {Adam}}, \ and\ \bibinfo
  {author} {\bibnamefont {{etal}}},\ }\href {\doibase
  10.1140/epjc/s10052-013-2617-1} {\bibfield  {journal} {\bibinfo  {journal}
  {The European Physical Journal C}\ }\textbf {\bibinfo {volume} {73}},\
  \bibinfo {pages} {2617} (\bibinfo {year} {2013})}\BibitemShut {NoStop}%
\bibitem [{\citenamefont {Aid}\ \emph {et~al.}(1996)\citenamefont {Aid},
  \citenamefont {Andreev}, \citenamefont {Andrieu},\ and\ \citenamefont
  {etc...}}]{AID19963}%
  \BibitemOpen
  \bibfield  {author} {\bibinfo {author} {\bibfnamefont {S.}~\bibnamefont
  {Aid}}, \bibinfo {author} {\bibfnamefont {V.}~\bibnamefont {Andreev}},
  \bibinfo {author} {\bibfnamefont {B.}~\bibnamefont {Andrieu}}, \ and\
  \bibinfo {author} {\bibnamefont {etc...}},\ }\href {\doibase
  http://dx.doi.org/10.1016/0550-3213(96)00274-X} {\bibfield  {journal}
  {\bibinfo  {journal} {Nuclear Physics B}\ }\textbf {\bibinfo {volume}
  {472}},\ \bibinfo {pages} {3 } (\bibinfo {year} {1996})}\BibitemShut
  {NoStop}%
\bibitem [{\citenamefont {Jones}\ \emph {et~al.}(2013)\citenamefont {Jones},
  \citenamefont {Martin}, \citenamefont {Ryskin},\ and\ \citenamefont
  {Teubner}}]{Jones2013}%
  \BibitemOpen
  \bibfield  {author} {\bibinfo {author} {\bibfnamefont {S.~P.}\ \bibnamefont
  {Jones}}, \bibinfo {author} {\bibfnamefont {A.~D.}\ \bibnamefont {Martin}},
  \bibinfo {author} {\bibfnamefont {M.~G.}\ \bibnamefont {Ryskin}}, \ and\
  \bibinfo {author} {\bibfnamefont {T.}~\bibnamefont {Teubner}},\ }\href
  {\doibase 10.1007/JHEP11(2013)085} {\bibfield  {journal} {\bibinfo  {journal}
  {Journal of High Energy Physics}\ }\textbf {\bibinfo {volume} {2013}},\
  \bibinfo {pages} {85} (\bibinfo {year} {2013})}\BibitemShut {NoStop}%
\bibitem [{\citenamefont {Griffioen}\ \emph {et~al.}(2015)\citenamefont
  {Griffioen}, \citenamefont {Arrington}, \citenamefont {Christy},
  \citenamefont {Ent}, \citenamefont {Kalantarians}, \citenamefont {Keppel},
  \citenamefont {Kuhn}, \citenamefont {Melnitchouk}, \citenamefont {Niculescu},
  \citenamefont {Niculescu}, \citenamefont {Tkachenko},\ and\ \citenamefont
  {Zhang}}]{PhysRevC.92.015211}%
  \BibitemOpen
  \bibfield  {author} {\bibinfo {author} {\bibfnamefont {K.~A.}\ \bibnamefont
  {Griffioen}}, \bibinfo {author} {\bibfnamefont {J.}~\bibnamefont
  {Arrington}}, \bibinfo {author} {\bibfnamefont {M.~E.}\ \bibnamefont
  {Christy}}, \bibinfo {author} {\bibfnamefont {R.}~\bibnamefont {Ent}},
  \bibinfo {author} {\bibfnamefont {N.}~\bibnamefont {Kalantarians}}, \bibinfo
  {author} {\bibfnamefont {C.~E.}\ \bibnamefont {Keppel}}, \bibinfo {author}
  {\bibfnamefont {S.~E.}\ \bibnamefont {Kuhn}}, \bibinfo {author}
  {\bibfnamefont {W.}~\bibnamefont {Melnitchouk}}, \bibinfo {author}
  {\bibfnamefont {G.}~\bibnamefont {Niculescu}}, \bibinfo {author}
  {\bibfnamefont {I.}~\bibnamefont {Niculescu}}, \bibinfo {author}
  {\bibfnamefont {S.}~\bibnamefont {Tkachenko}}, \ and\ \bibinfo {author}
  {\bibfnamefont {J.}~\bibnamefont {Zhang}},\ }\href {\doibase
  10.1103/PhysRevC.92.015211} {\bibfield  {journal} {\bibinfo  {journal} {Phys.
  Rev. C}\ }\textbf {\bibinfo {volume} {92}},\ \bibinfo {pages} {015211}
  (\bibinfo {year} {2015})}\BibitemShut {NoStop}%
\end{thebibliography}
\end{document}